\documentclass[aps,prb,twocolumn,float,amsmath]{revtex4-1}
\usepackage[dvips]{color}\usepackage{hyperref}
\usepackage[normalem]{ulem} 

\definecolor{g-blue}{rgb}{0.83,0.95,1}
\definecolor{g-yellow}{rgb}{1,1,0.7}

\definecolor{g-green}{rgb}{0.9,1,0.9}
\definecolor{green}{rgb}{0,0.6,0}
\definecolor{cyan}{rgb}{0,0.7,0.7}
\definecolor{black}{rgb}{0,0,0}
\definecolor{grey}{rgb}{0.4 ,0.4 ,0.4 }

\usepackage{bm}
\usepackage{amssymb}
\usepackage{amsfonts}
\usepackage{amsmath}
 \usepackage{graphicx}
  \usepackage{amsmath,bm,epsfig}

\def \ed {\end{document}}
\def\Fbox#1{\vskip1ex\hbox to 8.5cm{\hfil\fboxsep0.3cm\fbox{%
  \parbox{8.0cm}{#1}}\hfil}\vskip1ex\noindent}  

\newcommand{\eq}[1]{(\ref{#1})}
\newcommand{\Eq}[1]{Eq.\,(\ref{#1})}
\newcommand{\Eqs}[1]{Eqs.\,(\ref{#1})}
\newcommand{\Fig}[1]{Fig.\,\ref{#1}}
\newcommand{\Figs}[1]{Figs.\,\ref{#1}}
\newcommand{\Sec}[1]{Sec.\,\ref{#1}}
\newcommand{\Secs}[1]{Secs.\,\ref{#1}}
\newcommand{\Ref}[1]{Ref.\,\cite{#1}}
\newcommand{\Refs}[1]{Refs.\,\cite{#1}}

\def\be{\begin{equation}}\def\ee{\end{equation}}
\def\bea{\begin{eqnarray}}\def\eea{\end{eqnarray}}
\def\bse{\begin{subequations}}\def\ese{\end{subequations}}
\newcommand{\BE}[1]{\begin{equation}\label{#1}}
\newcommand{\BEA}[1]{\begin{eqnarray}\label{#1}}
\newcommand{\BSE}[1]{\begin{subequations}\label{#1}}

\let \nn  \nonumber  \newcommand{\br}{\\ \nn}

\let \= \equiv \let\*\cdot \let\~\widetilde \let\^\widehat \let\-\overline
\let\p\partial

  \def\1{\bm1} 

\def\<{\left\langle}    \def\>{\right\rangle}
\def\({\left(}          \def\){\right)}
 \def \[ {\left [} \def \] {\right ]}

\renewcommand{\a}{\alpha}\renewcommand{\b}{\beta}\newcommand{\g}{\gamma}

\renewcommand{\d}{\delta}
\newcommand{\ve}{\varepsilon}

\newcommand{\B}[1]{{\bm{#1}}}
 \newcommand{\C}[1]{{\mathcal{#1}}}    

\renewcommand{\sb}[1]{_{\text {#1}}}  
\renewcommand{\sp}[1]{^{\text {#1}}}  
\def\Sb#1{_{\scriptscriptstyle\rm{#1}}}
\def\He4 {$^4$He~}

\begin{document}

\title{Theory of energy spectra in superfluid  He-4 counterflow  turbulence }

\author{Victor. S. L'vov and  Anna Pomyalov }

\affiliation{ Department of Chemical and Biological Physics,  Weizmann Institute of Science,  Rehovot 76100,  Israel}

\begin{abstract}

In the thermally driven superfluid \He4 turbulence, the counterflow velocity $U\sb{ns}$  partially decouples normal and superfluid turbulent velocities. Recently we suggested [J. Low Temp. Phys.  \textbf{187}, 497 (2017)] that  this decoupling should tremendously increase the turbulent energy dissipation by mutual friction and significantly suppress the energy spectra. Comprehensive   measurements of the  apparent scaling exponent $n\sb{exp}$ of the 2$\sp{nd}$-order normal fluid velocity structure function  $S_2(r)\propto r^{n\sb{exp}}$ in the counterflow turbulence [Phys.Rev.B \textbf{96}, 094511 (2017)]  confirmed our scenario of gradual dependence  of the turbulence statistics on flow parameters. 
We develop an analytical  theory of the counterflow turbulence,  accounting for a two-fold mechanism of this phenomenon:
  i) a scale-dependent  competition between the turbulent velocity coupling by  mutual friction and the $U\sb{ns}$-induced turbulent velocity decoupling and ii) the turbulent energy dissipation by mutual friction enhanced by the velocity decoupling.
The suggested theory predicts the  energy spectra for a wide range of flow parameters.  The mean exponents of the normal fluid energy spectra $\<m\>\sb{10}$, found without fitting parameters, qualitatively agree with the  observed $n\sb{exp}+1$ for $T\gtrsim 1.85\,$K.      

\end{abstract}
\maketitle

 
 \subsection*{Introduction}
 Below  Bose-Einstein condensation temperature $T_\lambda\approx 2.17\,$K,
 liquid \He4 becomes a quantum inviscid superfluid\,\cite{Donnely,DB98,2}. The vorticity in superfluid \He4 is constrained to vortex-line singularities of
 core radius $a_0\approx  10^{-8}\,$cm and  fixed circulation $\kappa= h/M$,
 where $h$ is Planck's constant  and $M$ is the mass of the \He4
 atom\cite{Feynman}. The superfluid turbulence takes form of a
 complex tangle  of these vortex lines, with a typical inter-vortex distance\cite{Vinen}   $\ell\sim
 10^{-4}- 10^{-2}\,$cm.
 
 Large-scale hydrodynamics of such system can be described by a two fluid model,
 interpreting \He4  as a mixture of two coupled  fluid components:  a superfluid with
 zero viscosity and a viscous normal fluid. The contributions of the components to
 the mixture are defined by their densities $\rho\sb s, \rho\sb n: \rho\sb s+
 \rho\sb n=\rho$. Here $\rho$ is the density of $^4$He. The components  are
 coupled by a mutual friction force, mediated by the tangle of quantum
 vortexes\,\cite{Donnely,Vinen,Vinen2,Vinen3,37}.

 There is a growing consensus\cite{SS-2012,TenChapters,BLR} that large-scale
 turbulence in mechanically driven superfluid \He4  is similar to classical ``Kolmogorov"
 turbulence. In this case, both components move in the same direction and the mutual
 friction force couples them almost at all scales.  In this ``coflowing" quasi-classical superfluid \He4 Kolmogorov turbulence, the energy is supplied to the turbulent velocity fluctuations   by large-scale instabilities, and  is dissipated at small scales  (below so-called Kolmogorov microscale)  by viscous friction. Perhaps its  most important property is a step-by step energy transfer over scales with a constant ($k$-independent)  energy flux $\ve(k)=$const.  in the intermediate (or  ``inertial") interval of scales.

The superfluid turbulence, called ultra-quantum or Vinen's turbulence, may be excited  in $^4$He directly at scales of the order of $\ell$, for example,  by short pulses of electron beam\cite{Golov}.  In this case, there is no large-scale fluid motion. The tangle energy is dissipated by the mutual friction in the processes of vortex reconnections. During reconnection, sharp vortex tip causes very fast motion the vortex lines, that cannot be followed by the normal-fluid component due to its large inertia. The energy spectrum of  of such a turbulence  have a form of the peak with a maximum around $k\sb{max} \sim \pi/\ell$.  Here,  the energy is pumped and dissipated at the same scale and there is no  energy flux over scales: $\ve(k)=0$. 
 
In discussions of the energy spectra of superfluid turbulence in $^4$He, the  ``Kolmogorov turbulence" is often contrasted with the ``ultra-quantum" turbulence as two only forms of the energy transfer in the superfluids. However, we  note that   hydrodynamic turbulence in superfluid $^3$He, mechanically driven at large scales, can be considered as a third type of superfluid turbulence.  In this case, the normal-fluid component can be considered laminar (or resting) due to its very large kinematic viscosity. The energy cascade toward small scales is accompanied by energy dissipation at all scales caused by the mutual friction. Therefore the energy flux $\ve(k)$ is not constant, as in the Kolmogorov turbulence, and is not zero, as in Vinen's turbulence, but is a decreasing function of $k$.

  There is  one more, unique  way to generate turbulence in
 superfluid \He4  in a channel. When a heater is located at a closed end of a
 channel, while another end is open to a superfluid helium bath, the heat flux is
 carried away from the heater by the normal fluid alone.  To conserve mass, a superfluid
 current arises in the opposite direction. Here both components move relative to
 the channel walls with respective mean velocities $\B U \sb n$ and  $\B U\sb s$.
 In this way  a counterflow velocity $\B U\sb {ns}=\B U \sb n - \B U\sb s\ne 0$,
 proportional to the applied heat flux, is created along the channel, giving rise
 to a tangle of vortex lines.

 Systematic studies of counterflow turbulence have more than half of a century
 history, going back to classical 1957-papers of Vinen\cite{Vinen3}. Due to
 experimental limitations these studies were  mostly concentrated on global
 characteristics of the superfluid tubulence, such as $U\sb{ns}$-dependence of the intervortex
 distance, $\ell(U\sb{ns})$, [cf. reviews \cite{SS-2012,BLR}], the time evolution of the vortex-line-density\cite{Golov1,Golov2,Vinen4} and similar. The statistics of turbulent
 fluctuations was inaccessible. Only few years ago, with the development of
 breakthrough experimental visualization techniques,  the studies of the turbulent statistics of the normal\cite{WG-2015,WG-17a,WG-17b} 
  and superfluid\cite{Prague1,Prague2} components in the
 $^4$He counterflow become possible.
 
In particular,
 using thin lines of the 
 triplet-state He$_2$ molecular tracers created by a femptosecond-laser
 field ionization of He atoms\,\cite{18,WG-2015,WG-17a},
 one can measure the streamwise normal velocity across a channel,
 $v_x(y,t)$
 and  extract the transversal 2$\sp {nd}$-order structure functions
 \begin{equation*}
 S_2 (r)= \< |\delta_r   v_x(y,t)|^2 \>\,, \  \delta_r   v_x(y,t) \= v_x(y+r,t)-v_x(y,t)\,, 
 \end{equation*}
 of the  velocity differences $\delta_r$.   Here $\< \dots \>$ is a ``proper" averaging: over $y$, ensemble of visualization
 pulses  and  time (in the stationary regime) or over ensemble at fixed time
 delay after switching off the heat flux.  The physical meaning of $S_2(r)$ is
 the kinetic energy of turbulent velocity fluctuations (eddies, for
 shortness) of a scale $r$. For example, $S_2(r)\propto r^n$ means that the energy
 of eddies of size $r$ scales as $r^n$.
 
 Another way to characterize the energy distribution in the one-dimensional (1D)
 wavenumber $k$-space  is 1D energy spectrum $E(k, t)$, normalized such that the
 energy density per unit mass is defined as
 \begin{equation}\label{norm}
 E(t) = \frac{1}{2V} \int \< |\B u(\B r,t)|^2\> d\B r  = \int\limits _0 ^ \infty
 E(k,t) dk\, ,
 \end{equation}
 where $V= \int d \B r$ is the system volume.
 
 {In the scale-invariant situation, such as the inertial interval of scales in the 
 	classical hydrodynamic isotropic turbulence,  $E(k)\propto k^{-m}$. In this case, $S_2(r)$
 	may be reconstructed from  $E(k)$ (up to irrelevant for us dimensionless
 	prefactor)  as follows:
 	\begin{subequations}\label{rel} \begin{equation}\label{relA}
 		S_2(r)= \int\limits _0^\infty E(k)\Big [ 1- \frac {\sin (k\, r)}{k\, r}\Big ] dk\
 		.
 		\end{equation}
 		If $E(k)\propto k^{-m}$ and $m$ belongs to a so-called ``window of locality"\cite{LP-95}
 		$1 < m < 3$,
 		the integral\,\eq{relA} converges  and exponents $n$ and $m$ are related:
 		\begin{equation}\label{relC}
 		n=m-1\ .
 		\end{equation}\end{subequations}
 	For example,  Kolmogorov-1941 (K41) dimensional reasoning gives  $m\Sb{K41}=\frac
 	53$ (falls within the window of locality)  and simultaneously $n\Sb{K41}=\frac 23$, in
 	agreement with \Eq{relC}.

 	First measurements\cite{WG-2015} of $S_2(r)$ in the \He4 counterflow at
 	$T=1.83\,$K found that $S_2(r)\propto r$ (i.e. $n=1$) instead of  its K41 value
 	$n\Sb {K41}=\frac 23$:
 	\begin{subequations}\label{K41-a}
 		\begin{equation}\label{S2}
 		\mbox{K41:}\ S_2(r)\propto r^{2/3}\  \Rightarrow\  ^4\mbox{He:}\ S_2(r)\propto r
 		\ .
 		\end{equation}
 		Using the relation \Eq{relC}, the  normal-fluid energy spectrum was reconstructed in \Ref{WG-2015} as
 		\begin{equation}\label{E}
 		^4\mbox{He:}\ E\Sb{He4}(k)\propto k^{-2}\ .
 		\end{equation} \end{subequations}
 	
 	Observations\,\eqref{K41-a}, with an integer scaling exponent, stimulated attempts
 	to clarify a possible ``simple" underlying physical mechanisms. For example,
 	based
 	on the similarity of the spectrum \eqref{E} with the Kadomtsev-Petviashvili
 	spectrum\,\cite{KP} of the energy spectrum of strong acoustic turbulence,
 	$E(k)\propto k^{-2}$,
 	(cf. Online lecture course\,\cite{FNP}) one may think that the \He4
 	spectrum\,\eqref{E}  originates from (possible) discontinuities of the normal and
 	superfluid velocities   $v_x$ at planes, orthogonal to the counterflow direction
 	$\B {\hat x}$. Indeed, in the presence of randomly distributed velocity
 	discontinuities, their contribution to the $S_2(r)$ is proportional to their
 	number  between two space points, separated by $r$, i.e. $S_2(r)\propto r$, as
 	reported in \Ref{WG-2015}.
 	
 	However, to the best of our knowledge,  no analytical reasons or numerical
 	justifications  for these discontinuities were not found so far.  On the  contrary,
 	the developed analytic approach\cite{decoupling,BLPSV-2016,LP-2017} to the problem
 	of turbulent statistics of \He4 counterflow suggested a different scenario of this
 	phenomenon. It  was shown\cite{decoupling}  that in the counterflow turbulence, the
 	normal and superfluid turbulent velocity fluctuations $u\sb n$ and $u\sb s$
 	become  increasingly statistically independent (decoupled) as  their scale
 	decreases.
 	This decoupling  is due to sweeping of the normal fluid eddies by the
 	mean normal fluid velocity $U\sb n$, while the superfluid eddies are swept by the  mean
 	superfluid velocity $U\sb s$ in the opposite direction.  Therefore small scale
 	normal fluid and superfluid eddies do not have enough time to be correlated by
 	the mutual friction. This counterflow-induced decoupling  significantly increases  the energy dissipation
 	by mutual friction, leading\cite{BLPSV-2016,LP-2017} to a dependence   of the
 	turbulent statistics on the counterflow velocity. To what respect the  scenario\cite{LP-2017} reflects some (if any) aspects of
 	the turbulent statistics in counterflow was an open question.
 	
 	Recently,  systematic  experimental studies\cite{WG-17a} of the counterflow
 	turbulence statistics  in a wide range of temperatures $T$ and counterflow velocities
 	$U\sb{ns}$ were carried out. The normal velocity structure functions $S_2(r)$ were found to have
 	scaling behavior  $S_2(r)\propto r^n$ in an interval of scales about one decade with an apparent scaling exponent that
 	depends on both $U\sb{ns}$ and $T$, varying from about 0.9 to 1.4.
 
 	A first qualitative attempt  to understand theoretically the underlying physics of these new results, was
 	undertaken already in \Ref{WG-17a}. Main physical ideas, used  in this
 	approach,  largely overlap with those of the Weizmann group\cite{decoupling,LNV,BLPP-2012,BKLPPS-2017,LP-2017}, but are developed differently.

 	In this paper we offer a semi-quantitative theory of a stationary, space-homogeneous isotropic counterflow turbulence
 	 in superfluid $^4$He  for a wide range of temperatures $T$ and counterflow
 	velocities $U\sb{ns}$.   The theory  clarifies the details of complicated interplay
 	between competing mechanisms of  the  turbulent velocity coupling by  mutual friction and the $U\sb{ns}$-induced turbulent velocity decoupling, which, in addition, facilitates the turbulent energy dissipation by the mutual friction. Our main results are  the  turbulent energy spectra $E\sb n(k)$ and $E\sb s(k)$ of the normal and superfluid components of $^4$He in the wide range of the governing parameters: the temperature, the counterflow velocity, the vortex line density and Reynolds numbers.   In particular, we demonstrate that  the counterflow turbulence in $^4$He can be considered as the  most general form of superfluid turbulence that manifests characteristic features of all three types of turbulence, discussed above: \\
 (i) the quasi-classical Kolomogov-like  turbulence with a constant energy flux at   scales $r$ that exceed some cross-over scale $r_\times$, with  both fluid components well coupled by mutual friction;\\
  (ii) the $^3$He-like   turbulence  at scales $\ell <r< r_\times$, at which the  normal- and superfluid components become decoupled and the turbulent energy is dissipated by the mutual friction during energy cascade toward small scales, similar to $^3$He turbulence with decreasing energy flux;\\
 (iii)  the  ultra-quantum Vinen's turbulence with the energy spectrum peak at the intervortex scale $\ell$ and no energy flux.

 	The paper is organized as follows:
 	
 	In \Sec{s:theoryAll} we develop our analytical theory of the energy spectra in counterflow superfluid $^4$He turbulence.
 	Our approach is based on the coarse-grained Hall-Vinen-Bekarevich-Khalatnikov\cite{HV,BK}  equations of motion for the normal and superfluid turbulent velocities,  summarized in \Sec{ss:HVBK}. In \Sec{ss:theory} we derive the  balance equations for the normal and superfluid turbulent energy spectra $  E\sb n(k)$ and $  E\sb s(k)$. In \Sec{sss:int} we suggest an important improvement to the algebraic closure:  the self-consistent differential closure, that connects the energy fluxes over scales, $\ve\sb n(k)$ and $\ve\sb s(k)$    with the corresponding energy spectra $  E\sb n(k)$ and $  E\sb s(k)$ and their $k$-derivatives. The crucial component of the theory, the cross-correlation function of the normal and superfluid velocities\cite{decoupling}, is introduced in \Sec{ss:cross}.
 	In  the following \Secs{ss:middle} and \ref{sss:dimen} we formulate a simplified dimensionless version of the energy balance \Eq{20}, which is central to our current approach.

 	We present the results of the numerical solution of \Eq{20} in a wide range of parameters ($T, U\sb{ns}, \C L$, Re) in \Sec{s:ESCT} and analyze them in details in \Secs{ss:QSB}--\ref{ss:Re}. This allows us to clarify   how three
 	underlying physical processes:
 	i) the turbulent  velocity coupling by the mutual friction, characterized by a frequency $\Omega\sb{ns}$ \eqref{Ens1D},
 	ii) the velocity decoupling by  counterflow velocity and
 	iii) the energy dissipation by  mutual friction,
 	are competing.  As a result of
 	this competition, $  E\sb n(k)$ and $  E\sb s(k)$  have complicated behavior (cf. \Fig{ff:1}). In particular, we show that while the spectra are suppressed
 	compared to the classical K41-spectrum at all scales, the degree of this
 	suppression is scale-dependent: at small scales, the counterflow spectrum is less
 	suppressed for larger $\Omega\sb{ns}$, while at larger scales
 	the suppression becomes stronger with increasing $\Omega\sb{ns}$. The crossover scale, $k_{\times}$, depends
 	on both $\Omega\sb{ns}$ and $U\sb{ns}$ such that the resulting spectra are not scale-invariant (cf. Figs.\ref{ff:2} and \ref{ff:2b}).
 	Note that all results, discussed above, are related to the quasi-classical energy spectra of turbulent motion with the  scales $r$ much larger than the  intervortex  distance $\ell$. In \Sec{ss:QP} we consider the presumed quantum peak in the superfluid energy spectra, clarifying its intensity with respect to the  quasi-classical part of the superfluid spectra.  Our important result is that the quasi-classical and quantum parts of the superfluid spectra are well separated in the wave-number space, see \Fig{ff:6}.
 	
    In  \Sec{s:details} we apply out theory to the  range of parameters, similar to those realized in \Ref{WG-17a}. To this end, we first discuss and estimate in \Sec{ss:exp-par} the  dimensionless parameters ${\rm Re}_j, \~\Omega\sb{ns}$ and $\~U\sb{ns}$ that determine, according to our theory,  the energy spectra. These parameters for 11 experimental conditions at four temperatures $T=1.65\,\ 1.85,\ 2.0\,$K and  $T=2.1\,$K, are collected in Tab.\ref{t:2}.  The resulting pairs of  normal- and superfluid energy spectra are shown in \Fig{ff:4}. The relation between experimental apparent scaling exponents $n\sb{exp}$ of the 2$\sp{nd}$-order structure and the theoretical apparent scaling exponents of the energy spectra are discussed in \Sec{ss:app}.
 
 	Finally, we summarize our findings. We discuss  the restrictions and simplifications, used  in our  theory, and  delineate possible directions of further development.
 	In particular, we connect the discrepancy between  theoretical and experimental scaling exponents at low temperatures with the possible influence of the space inhomogeneity of the counterflow turbulence in a channel at low Reynolds numbers, which is not accounted for in our theory.

 	\section{\label{s:theoryAll} Energy balanced eqution}
 	
 	\subsection{\label{ss:HVBK} Gradually damped HVBK-equations for   counterflow
 		$^4$He turbulence}
 	
 	Following \Ref{DNS-He4,LP-2017,BKLPPS-2017}, we describe the large scale turbulence in
 	superfluid $^4$He   by the gradually-damped version\cite{He4} of the
 	coarse-grained  Hall-Vinen \cite{HV}-Bekarevich-Khalatnikov\cite{BK} (HVBK)
 	equations, generalized in \Ref{decoupling} for the counterflow turbulence.  In these equations the superfluid vorticity is assumed continuous, limiting their applicability to large scales with characteristic scale of turbulent fluctuations $R
 		> \ell $. HVBK equations  have a
 	form of two Navier-Stokes equations  for the turbulent velocity fluctuations $\B
 	u\sb n(\B r,t)$ and $\B u\sb s(\B r,t)$:
 	\begin{subequations}\label{NSE} \begin{eqnarray}   \label{NSEs} 
 		&& \hskip -1.3cm \frac{\p \,\B u\sb s}{\p t}+  [(\B u\sb s+\B U\sb s)\*
 		\B\nabla] \B u\sb s
 		- \frac 1{\rho\sb s }\B \nabla p\sb s  =\nu\sb s\,  \Delta \B u\sb s   + \B f
 		\sb {ns} \,, 
 		\\  \nn
 		&& \hskip -1.3cm \frac{\p \,\B u\sb n}{\p t}+[(\B u\sb n+\B U\sb n) \* \B
 		\nabla]\B u\sb n
 		- \frac 1{\rho\sb n }\B \nabla p\sb n = \nu\sb n\,  \Delta \B u\sb n
 		-\frac{\rho\sb s}{\rho\sb n}\B f \sb {ns} \, ,\\ \nn
 		&& \hskip -1.3cm  p\sb n =\frac{\rho\sb n}{\rho }[p+\frac{\rho\sb s}2|\B u\sb
 		s-\B u\sb  n|^2]\, ,
 		p\sb s =  \frac{\rho\sb s}{\rho }[p-\frac{\rho\sb n}2|\B u\sb s-\B u\sb
 		n|^2]\, ,\\ \label{5b}
 		\B f\sb {ns}&\simeq& \Omega\sb s  \,(\B  u \sb n-\B  u \sb s ) \,,
 		\quad \Omega\sb s  =  \a (T)  \kappa \C L\,,
 		\end{eqnarray}\end{subequations} 
 	coupled by the mutual friction force $\B f\sb{ns}$ in the
 	form \eqref{5b}, suggested in \Ref{LNV}.   It involves the turbulent velocity fluctuations of the
 	normal- and superfluid components, the
 	temperature dependent dimensionless dissipative mutual friction parameter $\alpha(T)$ and the
 	superfluid   vorticity  $\kappa \C L$, defined by the vortex line density $\C L , \,  \ell=\C L^{-1/2}$.
 	
 	Other parameters entering \Eqs{NSEs} include the pressures  $p\sb n$,  $p\sb s$ of the normal and the
 	superfluid components,
 	the  total  density $\rho\equiv  \rho\sb s+\rho\sb n$ of \He4 and the kinematic
 	viscosity  of normal fluid component $\nu\sb n=\eta / \rho \sb n$ with $\eta$
 	being the dynamical viscosity\cite{DB98}  of normal \He4 component.
 	The dissipative term with the Vinen's effective superfluid viscosity  $\nu\sb s$\,\cite{Vinen}
 	was added in \Ref{He4} to account for the energy dissipation at the intervortex
 	scale $\ell$ due to vortex reconnections, the energy transfer to  Kelvin waves and similar effects.
 	
 	Generally speaking,   \Eqs{NSEs} involve also   contributions of a reactive
 	(dimensionless)   mutual friction  parameter $\alpha'$, that renormalizes  the
 	nonlinear terms. For example, in \Eq{NSEs} $ (\B u\sb s\* \B
 	\nabla) \B u\sb s  \Rightarrow  (1-\alpha')(\B u\sb s\* \B
 	\nabla) \B u\sb s $.  However,  in the studied range of temperatures $|\alpha'|
 	\lesssim 0.02 \ll 1$   and this   renormalization can be ignored. For
 	similar reasons we neglected all other $\alpha'$-related terms in \Eqs{NSE}.

 	\subsection{\label{ss:theory}General energy balance equations }
 	Our theory is based on the stationary balance equations for the 1D energy spectra
 	$E\sb n(k)$ and $E\sb s (k)$ of the normal and superfluid components, defined by \Eq{norm}.
 	
 	To derive these equations, \Eqs{NSEs} were Fourier transformed, multiplied by the complex conjugates of the corresponding velocities and properly averaged. The pressure terms were eliminated using the incompressibility conditions. Finally, the energy balance equations have a form:
 	\begin{equation}\label{balance1}  \frac{d \ve_j}{d k} = \Omega_j \big
 	[E\sb{ns}(k) - E _j (k) \big ] - 2\, \nu_j k^2 E_j(k)\ .
 	\end{equation}
 	Here we use   subscript ``$_j$" to denote either  superfluid or normal fluid  $j\in \{\mbox{s, n}\}$ and define  $\Omega\sb n = \Omega\sb s \rho\sb s/
 	\rho \sb n$. The normal-fluid-superfluid  cross-correlation function $E\sb{ns}(k)$ is
 	defined similarly to \Eq{norm}:
 	\begin{equation}\label{Ens}
 	E\sb{ns}= \frac 1{2 V} \int \< \B u\sb n (\B r,t)\cdot \B u\sb s (\B r,t)\> d
 	\B r
 	= \int \limits _0^\infty E\sb {ns}(k)\, dk\ .
 	\end{equation}

 	The energy transfer term $ d \ve_j/dk $  in \Eq{balance1} originates from the
 	nonlinear terms in the HVBK \Eqs{NSE}  and has the same form\cite{LP-95,LP-2,BL} as in classical
 	turbulence:
 	\begin{subequations}\label{Tr}
 		\begin{eqnarray}\nn
 		\frac{d \ve_j(k)}{dk}&=& 2\, \mbox{Re}\Big\{\int V^{\xi\beta\gamma}(\B k,\B q,\B
 		p)\, F_j^{\xi\beta\gamma}(\B k,\B q,\B p)\\ \label{genA}
 		&& \times \delta(\B k+\B q+\B p)\frac{d^3 q \, d^3 p}{(2\pi)^6} \, \Big\}\,, \\
 		\nn
 		V^{\xi\beta\gamma}(\B k,\B q,\B p)&=& i  \Big ( \delta _{\xi \xi'}- \frac{ k^\xi
 			k^{\xi'}}{k^2}   \Big )\\ \label{genB}
 		&& \times \Big( k^\beta \delta_{\xi ' \gamma} + k^\gamma \delta _{\xi' \beta}
 		\Big )\ .
 		\end{eqnarray}
 	\end{subequations}
 	Here $F_j^{\xi\beta\gamma}(\B k,\B q,\B p)$ is the  simultaneous
 	triple-correlation function of turbulent (normal or superfluid) velocity fluctuations
 	in the $\B k$-representation, that  we will not specify here and $
 	V^{\xi\beta\gamma}(\B k,\B q,\B p)$ is the interaction vertex in the HVBK (as
 	well as in the Navier-Stokes) equations.
 	Importantly,  the right-hand-side of \Eq{genA} conserves the total
 	turbulent kinetic energy (i.e. the integral of $E_j(k)$ over entire $\B k$-space)
 	and therefore can be written in the divergent form  as $ {d \ve_j }/{d k}$.

 	\subsection{\label{sss:int} Self-consistent differential closure}
 	
 	One of the main problems in the theory of hydrodynamic turbulence is to find the
 	triple-correlation function $ F ^{\xi\beta\gamma}(\B k,\B q,\B p)$, which
 	determines the energy flux $\ve(k)$ in \Eqs{Tr}. The simplest way is to directly model
 	$\ve(k)$ using dimensional reasoning to connect $\ve(k)$ and the
 	energy spectrum $E(k)$ with the same wavenumber $k$:
 	\begin{equation}\label{K41}
 	\ve(k)= C k ^{5/2} E^{3/2}(k)\,,
 	\end{equation}
 	Here  $C$ is  the phenomenological constant with the  value $C\approx 0.5$, corresponding to the fully developed turbulence of classical fluid \cite{CK,YeungZhou}.  The algebraic
 	closure\,\eqref{K41} is  based\cite{Fri} mainly on the   Kolmogorov-1941  (K41) assumption
 	of the universality of turbulent statistics in the limit of large Reynolds numbers
 	and on  the Richardson-1922 step-by-step cascade picture of the energy transfer
 	towards large $k$. The energy-cascade picture combined with the K41  idea that in this case $\ve(k)$ is the
 	only relevant physical parameter determining  the level of  turbulent excitations and their statistics, lead to \Eq{K41}.
 	
 	More realistic modeling of $\ve(k)$ can be reached in the framework of integral
 	closures, widely used in analytic theories of classical turbulence\cite{Fri}, for
 	example, so-called   Eddy-damped quasinormal Markovian (EDQNM) closure  or
 	Kraichnan's Direct Interaction Approximation\,\cite{Kra1,Kra2}.  These closures are
 	based on the representation of  third-order velocity correlation function
 	$F^{\xi\beta\gamma}\sb{s}$ in \Eq{genA} as a product of the vertex $V$,
 	\Eq{genB},  two second-order correlations $E(k)$,  and  the response (Green's)
 	functions $G(k) \sim \Gamma(k)-$the typical relaxation frequencies at scale
 	$k$. Keeping in mind the uncontrolled character of integral closures,   L'vov,
 	Nazarenko and Rudenko (LNR) suggested in \Ref{LNR} a simplified version of EDQNM
 	closure with the same level of justification for isotropic turbulence. LNR
 	replaced  a volume element [$d^3 q
 	\, d^3 p \, \d^3( \B k +\B  q+ \B  p) $] in \Eq{genA}, involving 3-dimensional vectors
 	$\B  k$,   $\B  q$, and $\B p$, by its isotropic version [$ q^2 d q \, p^2 d   p
 	\  \d (  k +  q+   p)/ (k^2+q^2+p^2)$],  involving only
 	one-dimensional vectors  $  k$,   $  q$, and $  p$  varying
 	over the interval $(-\infty, + \infty)$. In addition, they replaced the interaction amplitude
 	$V^{\xi\b\g}(  \B k, \B q, \B p)$, \Eq{genB}  by its  scalar  version $ (i  k)$.
 	The resulting LNR closure can be written as follows:
 	\begin{subequations}\label{24}
 		\begin{eqnarray}\label{24A}
 		&&\frac{d  \ve(k)} {d k }=\frac { A_1\, k }{2\pi^2}
 		\int  _{ -\infty}^\infty
 		\frac{  d   q \,  d   p
 			\, \d (  k +  q+   p)}{2\pi \, (k^2+q^2+p^2)}\br
 		&&\hskip -.5cm   \times \frac{       k^3\, E(|q|)  E (|p|) +
 			q^3\,E(|k|) E (|p|) +    p^3 \,E(|q|) E(|k|)
 		}{\Gamma(|k|)+\Gamma(|q|)+\Gamma(|p|)} \  .
 		\end{eqnarray} 
 		Here $A_1$ is a dimensionless parameter of the order of unity.
 		
 		The integral closure\,\eqref{24A} may be related to the algebraic closure\,\eqref{K41} by assuming that the integral\,\eqref{24A} converges (i.e.  the
 		main contribution to it comes from the wavenumbers of similar scales $q\sim p \sim
 		k$, so-called locality of interaction) and  estimating $\Gamma(k)$ as $\sqrt {k^3 E(k)}$ and $d \ve (k)/ d k $ as
 		$\ve(k)/k$.

 		The LNR model~\eq{24A}  satisfies all the general closure requirements: i) it
 		conserves energy, $\int   \frac{d \ve(k)}{d k}dk =0$  for any  $E(k)$; ii)
 		$\frac{d \ve(k)}{d k} =0 $ for the thermodynamic equilibrium spectrum
 		$E(k)\propto k^2$ and  for the cascade K41 spectrum $ E(k)\propto |k|^{-5/3}$.
 		Importantly, the integrand in \Eq{24A}  has the correct  asymptotic
 		behavior in the limits of small and large $q/k$, as required by the
 		sweeping-free Belinicher-L'vov representation\,\cite{BL}. This
 		means that the model\,\eqref{24A} adequately  reflects the contributions of the
 		extended-interaction triads and thus  can be used for the analysis of the
 		non-local energy transfer, which become important\cite{BKLPPS-2017} when the scaling exponent $m$ approaches $m=3$.
 		
 		The mutual friction terms in \Eqs{NSEs} cause additional energy dissipation at all scales. Therefore, we can expect that the energy spectra may be  steeper than K41 and not scale-invariant.
 		
 		To generalize  the closure\,\eqref{24A} for such a situation, assume that
 		the energy spectrum has a form $E(k)=E_0 / k^{m}$ with 	a  scaling exponent $m \geqslant 5/3$.  As $m \to 3$, the main contribution to the  integral\,\eqref{24A} comes from the distant interactions with wavevectors of different scales.
 		In particular, the $\delta$-function in the integral dictates that for $q \ll  k$, $p\approx -
 		k$  and the integral\,\eqref{24A} may be approximated as:
 		\begin{eqnarray}\label{24B}
 		\frac{d  \ve(k)} {d k }&\approx & \frac { A_1\,  E_0 }{8\pi^3 k \Gamma(|k|)}
 		\int  _{ -k}^k
 		\frac  {d   q }{q^m} \br
 		&& \times \big [k^3 \, E(|k+q|)- (k+q)^3 E(k)\big ] \br
 		&& \hskip - 1.5cm \approx   \frac { A_1\,  E_0 }{8\pi^3 k \Gamma(|k|)} \,
 		\frac{d ^2}{dq^2} \big [k^3 \, E(|k+q|)- (k+q)^3 E(k)\big ]  _{q\to 0}\\
 		\label{24C}
 		&& \times  \int\limits  _0^\infty q^{2-m} dq\simeq \frac{A_1 \big[ k E(k) \big
 			]^{3/2}} {8\pi^3 (3-m)}\ .
 			\end{eqnarray}
 	\end{subequations}
 		Dimensionally, \Eq{24C} coincides with the K41 algebraic closure\,\eqref{K41}, but
 		contains additional prefactor $1/(3-m)$.  This may be interpreted as  $m$-dependent parameter
 		$C$ in front of \Eq{K41}, which diverges
 		as $m\to 3$. The physical reason for such a dependence is simple: as $m$ increases,
 		more and more extended triads that involve $k$- and $(q\ll k)$-modes contribute
 		to the energy influx in the $k$-mode. As a result,  the energy flux grows
 		$\propto 1/(3-m)$, according to \Eq{24C}. Moreover, when $m \geqslant  3$ the
 		integral\,\eq{24C} formally diverges, meaning that the leading contribution to the flux at $k$-mode
 		comes not from comparable $(q\sim k)$-modes (as assumed in the
 		Richardson-Kolmogorov step-by-step cascade picture of the energy flux), but
 		directly from the largest, energy-containing modes in the turbulent flow.
 		
 		Thus, by accounting for the possible scale-dependent nonlocal energy transfer over scales, we generalize the standard K41-closure\,\eqref{K41} by  including the $k$-dependence of the prefactor $C$:
 	\begin{subequations}\label{LP17}
 	        \begin{equation}\label{LP17a}
 		\ve(k)= C(k) k ^{5/2} E^{3/2}(k)\,, \  C(k)= \frac{4\, C}{3\,[3-m(k)]}\ .
 		\end{equation}
 		For convenience, the prefactor $C(k)$ is chosen to reproduce the Kolmogorov constant
 		$C$ for the K41 scaling exponent $m(k)=5/3$.  As follows from above
 		arguments, the function of $m(k)$ in \eqref{LP17a} should be
 		understood as a  local  scaling exponent of $E(k$):
 		\begin{equation}\label{LP17b}
 	         m(k)=\frac{d \ln E(k)}{d \ln (k)} \ ,
 		\end{equation}
 	\end{subequations}
                 making the new closure \emph{a self-consistent differential closure}. 
 	 
 		\subsection{\label{ss:cross} Cross-correlation function}
 		The general form of the cross-correlation function\cite{decoupling} $E\sb{ns}$  reads as:
 	\begin{subequations}\label{Ens1}
 		\begin{eqnarray}\label{Ens1A}
 		E\sb{ns}(k)&=&D(k)E\sb{ns}^{(0)}(k)\,, \ D( k)= \frac{\arctan
 			[\xi(k)]}{\xi(k)}\,, ~~~~~~~~\\
 		\ \xi(k)&=& \frac{k}{k_\times}\,, \quad  k_\times= \frac{\Gamma(k)}{U\sb{ns}}\ .
 		\end{eqnarray}
 		Here  $D(k)$ is the $U\sb {ns}$-dependent decoupling function,  and
 		$E\sb{ns}^{(0)}(k)$  has the form\cite{LNS,He4}:
 		\begin{eqnarray}
 		\label{Ens1C}
 		E\sb{ns}^{(0)}(k) &=& \frac{ \Omega\sb{ns}    \big [\rho\sb n E\sb n(k)+
 			\rho\sb s E\sb s(k)\big ]  }{ \Gamma(k) \, \rho   } \,, \\ \label{Ens1D}
 		\Gamma(k) &=& \Omega\sb{ns}   + \gamma\sb s(k)+ \gamma\sb n (k)+ (\nu\sb s+
 		\nu\sb n)\, k^2\,,\\ \nn
 		\Omega\sb{ns} &=&\Omega\sb n+ \Omega \sb s= \alpha \kappa \C
 		L\frac{\rho}{\rho\sb n}  \,,  \
 		\gamma_j(k)= C _\gamma \sqrt {k^3 E_j (k)} \ . ~~~
 		\end{eqnarray}
 	\end{subequations}
 	Here $C_\gamma$ is a phenomenological parameter, the same for both components.
 	However $E\sb n(k)$ and $E\sb s(k)$ in \Eq{Ens1C} are \emph{not} the energy spectra at
 	$U\sb{ns}=0$, but the $U\sb{ns}$-dependent energy spectra, found
 	self-consistently by solving \Eqs{balance1} with $E\sb {ns}(k)$  given by \Eqs{Ens1}.

 	\begin{table}
 		\begin{tabular}{||c|  c| c|c|c|c|c||}
 			\hline\hline
 			        $T$, K         & 1.4    & 1.65  & 1.85  & 1.95  &  2.0  &  2.1  \\ \hline
 			   $\rho\sb n/\rho$    & 0.0728 & 0.193 & 0.364 & 0.482 & 0.553 & 0.741 \\ \hline
 			       $\alpha$        & 0.051  & 0.111 & 0.181 & 0.236 & 0.279 & 0.481 \\
 			$\alpha\rho/\rho\sb n$ & 0.701  & 0.575 & 0.497 & 0.489 & 0.504 & 0.649 \\ \hline
 			 $  \nu\sb n/\kappa$   & 1.34   & 0.462 & 0.248 & 0.199 & 0.182 & 0.167 \\
 			        	$\nu\sb s '/\kappa$ & 0.135 &0.228  &0.265 & 0.296& 0.312 &0.427 \\
 			
 			\hline\hline
 			
 		\end{tabular}
 	
 	\caption{\label{t:1}The parameters\,\cite{DB98} of the superfluid $^4$He:
 			the  normal component fraction $\rho\sb n/\rho$;  the mutual
 			friction parameter $\alpha$, the combination
 			$\alpha\rho/\rho\sb n$ ;
 			the  kinematic viscosity of the normal-fluid
 			component $\nu\sb n\equiv \mu/\rho\sb n$;
 			the  effective superfluid viscosity\,\cite{Vinen} $\nu'\sb s$.}
 	\end{table}

 	\subsection{\label{ss:middle}  Simplified energy
 		balance equation}
 	
	The \Eqs{balance1} are coupled via cross-correlation function
 	$E\sb{ns}$, which depends on the energy spectra of both components.
 	To find the leading contributions to $E\sb{ns}$,  we recall that at $k$ close to
 	$k_0$  the velocities are well correlated\cite{decoupling}, meaning that $E\sb
 	n(k)\simeq E\sb s(k) $, while for $k\gg k_0$   they are almost decorrelated
 	and $E\sb{ns}(k)$ is negligible compared to $E\sb n$ and  $E\sb{s}$.
 	Therefore,  without loss of accuracy we can  replace $E\sb n(k)$ by $E\sb s(k)$
 	in  \Eqs{Ens1} for  $E\sb {ns}$  that enters into  balance equation  \eqref{balance1} for the superfluid
 	and $E\sb s(k)$ by $E\sb n(k)$  for
 	$E\sb {ns}$  that enters into  normal-fluid balance equation.
 	
 	In this way we obtain   decoupled equations for the fluxes $\varepsilon\sb n$ and
 	$\varepsilon\sb s$
 	\begin{subequations}\label{Balance1}
 		\begin{eqnarray}\label{Balance1A}
 		\frac{d \ve_j}{d k} &=& E_j(k) \big \{ \Omega   [  D_j(k)-1  ]    - 2\, \nu_j
 		k^2  \big \} \,,
 		\end{eqnarray}
 		with different decoupling functions for the normal-fluid and superfluid components:
 		\begin{eqnarray}\label{Balance1B}
 		D_j(k)&=& \frac {\Omega\sb {ns}\arctan  [k U\sb {ns}/ \Gamma_j(k)]}{k U\sb
 			{ns}} \,,  \end{eqnarray}
 		and different values of full damping frequencies:
 		\begin{eqnarray} \label{Balance1C}
 		\Gamma_j(k) &=& \Omega\sb {ns}+ 2 C_\gamma  \sqrt { k^3E_j(k)}+ (\nu\sb s +
 		\nu \sb n)k^2 \ .
 		\end{eqnarray}
 	\end{subequations}
 	The balance \Eqs{Balance1}, being uncoupled for the normal-fluid and
 	superfluid components,   are already much simpler than the fully coupled balance equations
 	\eqref{balance1} and \eqref{Ens1}.
 	
 	We are now ready to make next step and to analyze the relative importance of
 	different contributions to the damping frequencies $\Gamma_j(k)$,   comparing the
 	dissipation due to mutual friction  $\Omega \sb {ns}$, the rate of the energy
 	transfer between scales $C_{\gamma}\sqrt{k^3 E(k)}$ and the viscous dissipation
 	$(\nu\sb s +\nu\sb n)k^2$. We start with $\Omega \sb {ns}=\alpha (T)\kappa \C L
 	\rho/ \rho\sb n(T)$. In the intermediate range of temperatures  $\alpha (T)
 	\rho/ \rho\sb n(T)\approx 0.5$,  weakly depending  on $T$, cf Tab.\,\ref{t:1}.
 	Therefore we can peacefully take $\Omega \sb {ns}\approx 0.5 \kappa \C L$.
 	
 	Next,  the sum $(\nu\sb s+ \nu\sb n)$ in this temperature
 	interval also weakly depends on $T$ and is very close to $0.5\,\kappa$.  (cf. Tab.\,\ref{t:1} and
 	Fig.\,5 in \Ref{He4} and  \Ref{LSS} for discussion on physical origin of the numerical value). Estimating the largest wavenumber of the inertial
 	interval by the quantum cutoff $k\sb {max}\sim 1/\ell =\sqrt{\C L}$, we find
 	that $(\nu\sb s+ \nu\sb n)k\sb{max}^2\simeq 0.5 \kappa \C L \approx
 	\Omega\sb{ns}$.
 		Therefore,  $(\nu\sb s+ \nu\sb
 	n)k^2<  \Omega\sb{ns}$ in the entire inertial interval, except for its large-$k$
 	end, where the accurate representation of the decoupling functions is not important.
 	This allows us to neglect in most cases the viscous contributions in
 	\Eq{Balance1C} for $\Gamma_j(k)$.

 	To estimate the energy transfer terms, recall that in  the classical  K41
 	turbulence (with $E(k)\propto k^{-5/3}$) $\sqrt{k^3 E(k)}$ grows as $k^{2/3}$. It
 	remains larger than viscous terms in the entire inertial range and become
 	compatible with  $\nu k^2$ at the large-$k$ end of the inertial interval, at the
 	Kolmogorov microscale.  In our case, the spectra  $E_j(k)$ decay with $k$  faster than the K41
 	spectrum, owing to the energy dissipation due to
 	mutual friction.    Therefore, in the
 	inertial interval of counterflow turbulence   $\sqrt{k^3 E_j(k)}$ grows slower
 	than in K41 turbulence and  remains smaller than $\Omega\sb{ns}$ for all $k$.
 	Moreover, recent estimate of $C_\gamma$  using Direct Numerical Simulations of the superfluid $^4$He
 	turbulence\cite{DNS-He4} gives $C_\gamma\ll 1$. Therefore, for these conditions we can
 	neglect  the energy transfer terms compared to $\Omega\sb{ns}$.
 	Having all these in mind, we approximate $\Gamma_j$   as $\Omega\sb{ns}$ and,
 	using \Eq{Balance1B},  get
 	\begin{equation}\label{newD}
 	D\sb n(k)=D\sb s(k)\= D(k)\,,
 	\end{equation}
 	with $\xi(k)= k U\sb{ns} / \Omega\sb{ns}$.

 	Now, we combine   \Eqs{LP17}, \eq{Balance1A} and  \eqref{newD} to finalize
 	an approximation for the balance equations 
 	\begin{subequations}\label{fin-bal}
 		\begin{eqnarray}\nn
 		C(k) \, \frac d {dk} k^{5/2} E_j ^{3/2}(k)&=& E_j (k)\big \{ \Omega_j [ D(k)-1]\\
 		\label{fin-balA} && - 2 \nu_j k^2 \big \}  , ~~~~~~\\ \label{fin-balB}
 		D(k)= \frac{\arctan [\xi(k)]}{\xi(k)}\,, && \quad \xi(k)= \frac{k U\sb
 			{ns}}{\Omega\sb{ns}} \ .
 		\end{eqnarray} \end{subequations}
 	
 		In deriving \Eq{fin-balA} we neglected for simplicity the $k$-derivative of $m(k)$ in the expression for $C(k)$ with respect to $d[ k^{5/3}E(k)]/d k$ since $m(k)$ varies between 5/3 to 3 in the entire range of $k$, while $ k^{5/3}E(k) $ varies by many orders of magnitude.

 		\subsection{\label{sss:dimen} Dimensionless form  of  the energy balance equation}
 		To analyze the energy balance \Eq{fin-bal} and to open a way to its numerical solution,    we first 
 		rewrite  it  in the dimensionless form.  To this end we  introduce a dimensionless wavenumber $q$   and a
 		dimensionless energy spectrum $\C E (q)$:
 		\begin{equation}\label{nodim}
 		q=k/k_0 \,, \quad \C E (q)= E(k)/E(k_0)\ .
 		\end{equation}
 		Here the minimal wave number is estimated as $k_0=2\pi / \Delta$, where
 		$\Delta$ is outer scale of turbulence.
 		
 		The resulting dimensionless equations for new dimensionless functions
 		\begin{subequations}\label{psi}
 			\begin{equation}\label{psiA}
 			\Psi_j(q)= \sqrt {q^{5/3}\C  E_j(q)}\,,
 			\end{equation}
 			take the form
 			\begin{eqnarray}\label{psiB}
 			\frac{d \Psi_j(q)}{d q} &=&A_j\, \frac{ D(q k_0)-1   }{q^{5/3}} - a_j \,
 			q^{1/3}\,,~ j\in\{s,n\}\,,~~~~
 			\\ \label{psiC}
 			A_j&=&  \frac {\Omega_j}{3 C (k)  \sqrt {k_0^3  E_j  (k_0)}}\,,\\ \label{psiD}
 			\  a_j &=& \frac{2 \nu_j  \sqrt {k_0}}{3  C (k) \sqrt {E_j  (k_0)}}\ . ~~~~~
 			\end{eqnarray} 
 		\end{subequations}

 		To clarify further the parameters in \Eqs{psi}, we define the dimensionless parameters that govern the counterflow superfluid turbulence: the turbulent Reynolds numbers, the efficiency of dissipation by mutual friction and the dimensionless counterflow velocity.
 		
 		Similar to the classical hydrodynamic turbulence, the energy dissipation by
 		viscous friction in the superfluid turbulence is governed by the Reynolds number. There are two such
 		Reynolds numbers, ${\rm Re}\sb n $ and ${\rm Re}\sb s$: 
 		\begin{subequations}\label{dimen}
 		\begin{equation}\label{BalC}
 		{\rm Re\!}_j =\frac{u\Sb T}{k_0 \nu_j}\,, \quad 
 	  u\Sb T \simeq  \sqrt {k_0 E (k_0)}		 \ .
 		\end{equation}
 		Here  we ignore the presumably small difference  between velocity fluctuations of the components at  scale $k_0$ [i.e assume $E\sb s(k_0)=E\sb n(k_0)=E(k_0)$] and  estimate the root-mean-square (rms)   turbulent fluctuations $u\Sb T$ as $\sqrt {k_0 E (k_0)}$.    The 
 		ratio of the Reynolds numbers \eqref{BalC} is defined by the viscosities ${\rm Re}\sb s / {\rm Re}\sb n= \nu\sb n/ \nu \sb s$ and depends  only on the temperature. Therefore, for a given temperature we are
 		left with only one Reynolds number, say ${\rm Re}\sb n$.
 		
 		There is one more mechanism of the energy dissipation in superfluid turbulence: the dissipation by
 		mutual friction. This kind of dissipation is characterized by the frequency $\Omega\sb{ns}$, \Eq{Ens1C}.
 	As was mentioned above, $\Omega\sb{ns}\approx 0.5 \kappa \C L$ in the entire temperature range. Then, the
 		partial frequencies $\Omega\sb s$ and $\Omega\sb n$ that govern the energy
 		dissipation by mutual friction in the superfluid and normal-fluid components with a
 		given $\Omega\sb{ns}$, depend  only on the densities $\rho \sb s$ and $\rho
 		\sb n$, according to \Eq{Ens1C}. Therefore we can say that for a given temperature,
 		the dissipation by mutual friction  is governed by only one frequency $\Omega
 		\sb{ns}\propto \C L$.
 		
 		As was shown in \Refs{LNV,BLPP-2012,BKLPPS-2017}, the rate of energy dissipation by
 		mutual friction should be compared with the $k$-dependent rate of the energy
 		transfer over the cascade, characterized by the turnover frequency of the eddies of
 		scale $1/k$, $\gamma_j(k)$, \Eq{Ens1C}. This dictates a natural normalization of
 		$\Omega\sb{ns}$ by $\gamma (k_0)$, which can be estimated as $k_0 u\Sb T$. In
 		other words,  we suggest to use
 			\begin{equation}\label{Omega}
 			\~ \Omega\sb{ns} \= \Omega \sb {ns} / k_0 u\Sb T\,,
 			\end{equation}
 			as the dimensionless parameter that characterizes the efficiency of dissipation by mutual friction .
 			
 			Last important parameter of the problem is the counterflow velocity $U\sb{ns}$. It
 			is natural to normalize it by the turbulent velocity $u\Sb T$, introducing a
 			dimensionless velocity
 			\begin{equation}\label{Uns}
 			\~ U\sb{ns}\= U\sb {ns}/  u\Sb T\,,
 			\end{equation}
 		\end{subequations}

	Using parameters\,\eqref{dimen} together with \Eqs{LP17} and \eqref{fin-balB} we rewrite \Eq{psi} as follows
 	\begin{subequations}\label{20}
 		\begin{eqnarray}\label{20A}
 	 	4 C  \,\frac{d \Psi_j}{dq} &=&- \Big [ \frac23 + \frac{d \Psi_j}{dq} \, \frac{q}{\Psi_j}\Big] \\ \nn
 		&&\times \Big \{   \frac{\~\Omega_j}{q^{5/3}}   \Big[1 -\frac{\arctan(q/q_\times)}{q/q_\times}\Big] 
+ \frac{2}{\mbox{Re}_j}\Big\}\\ \label{20B}
\~ \Omega\sb n=\~ \Omega\sb{ns}\frac{\rho\sb s}{\rho}\,,&& \~\Omega\sb s= \~ \Omega\sb{ns}\frac{\rho\sb n}{\rho}\,, \quad 
q_\times= \frac{\~\Omega\sb {ns}}{\~ U\sb{ns}}\ .
 	\end{eqnarray}	 
 \end{subequations}
 These are the first-order ordinary differential equations for $\Psi_j(q)$. Aside from the temperature dependent parameter $\rho\sb n/\rho$, they include four dimensionless
 parameters that characterize the superfluid counterflow   turbulence: $\~ \Omega\sb{ns}$, $\~ U\sb{ns}$ and Re$_j$.

 	\begin{figure*}
 		\begin{tabular}{ccc}
 			$T=1.65\,$K, $ \ \~ U\sb {ns}=1.0 $& 	$T=1.65\,$K, \ $ \~ U\sb {ns}=4.0 $  &	$T=1.65\,$K,\  $ \~ U\sb {ns}=8.0 $ \\
 			\includegraphics[scale=0.45 ]{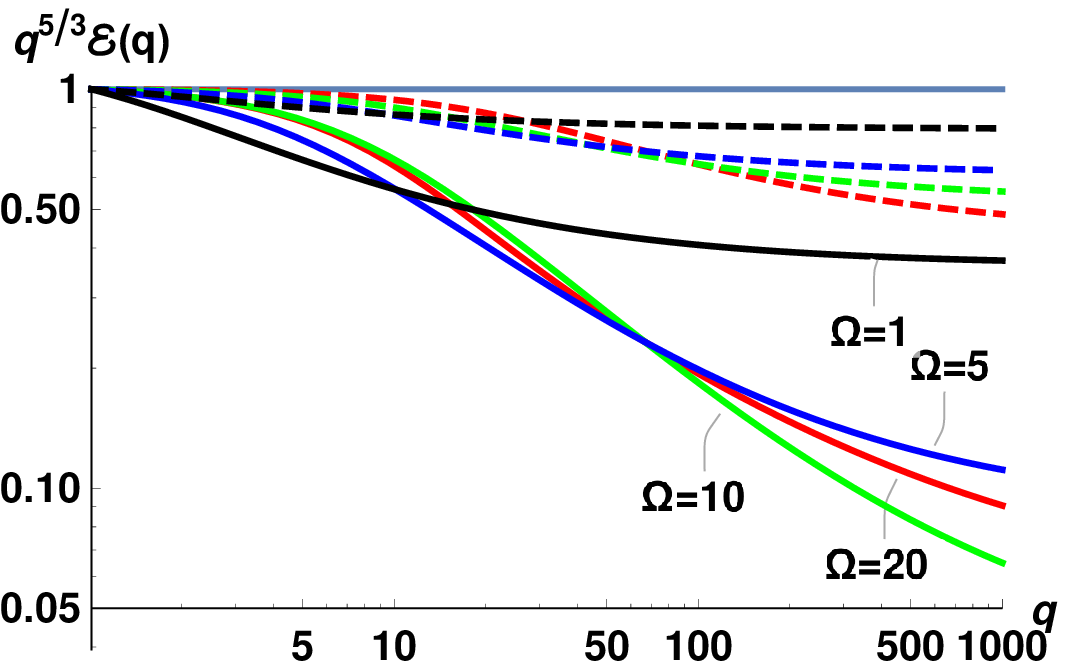}&
 			\includegraphics[scale=0.45 ]{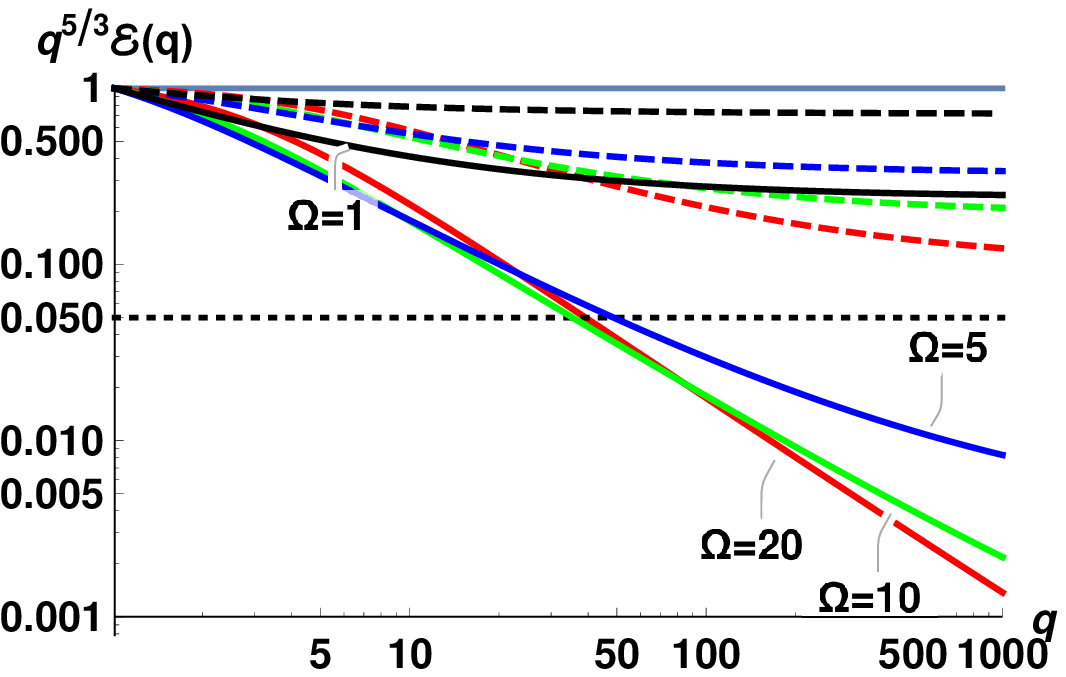}&
 			\includegraphics[scale=0.45 ]{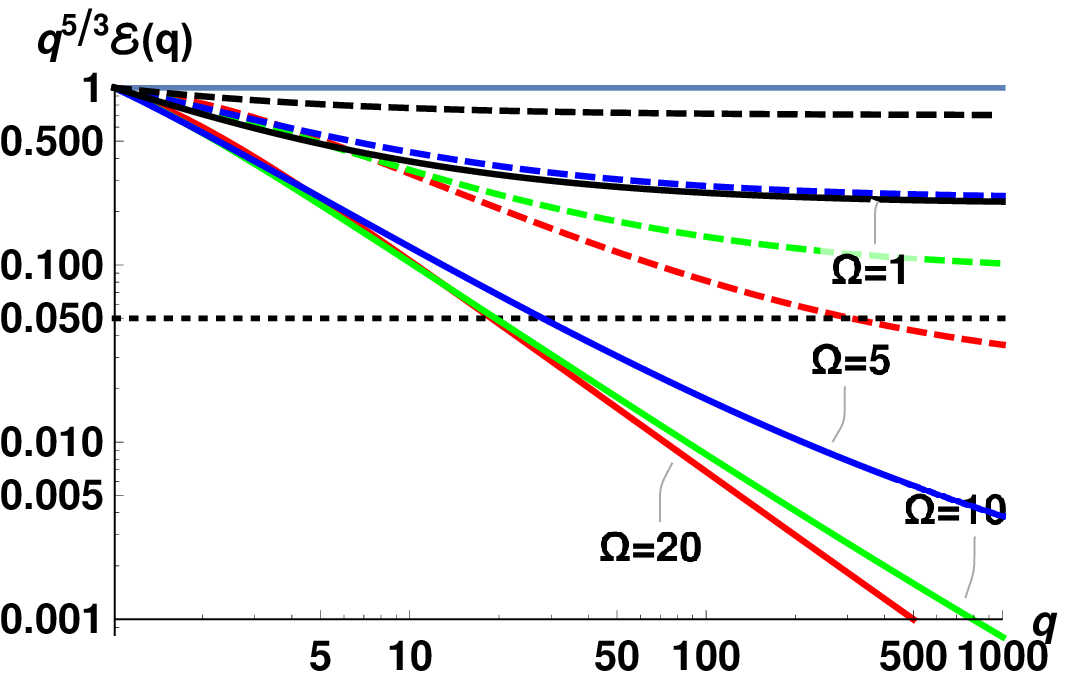}\\
 		$T=1.85\,$K, $ \ \~ U\sb {ns}=1.0 $& 	$T=1.85\,$K, \ $ \~ U\sb {ns}=4.0 $  &	$T=1.85\,$K,\  $ \~ U\sb {ns}=8.0 $ \\
 			\includegraphics[scale=0.45 ]{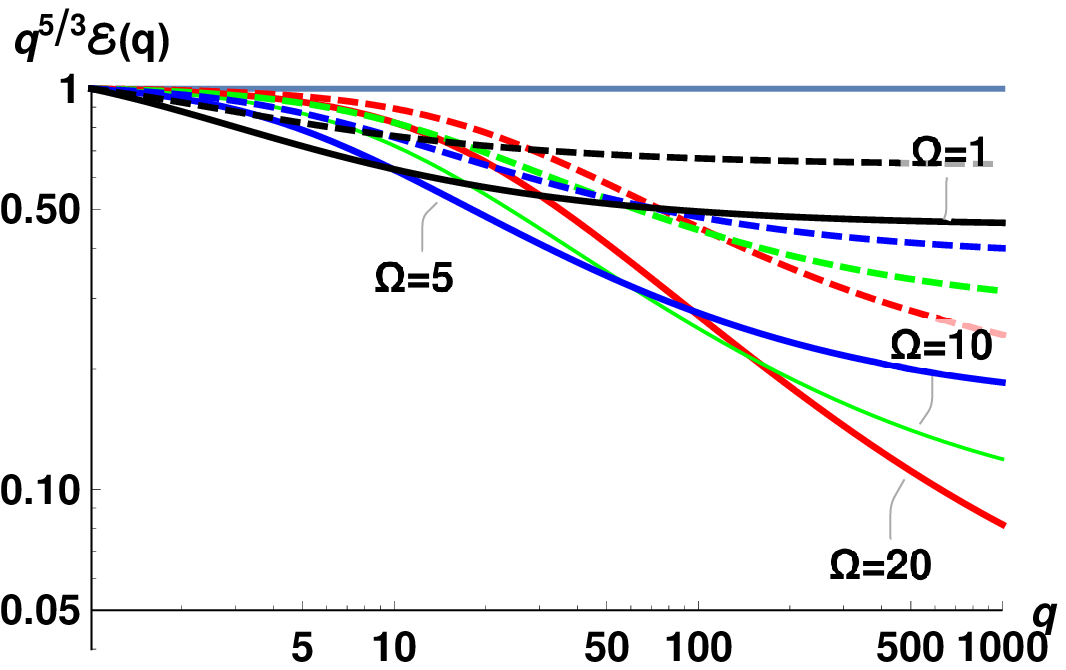}&
 			\includegraphics[scale=0.45 ]{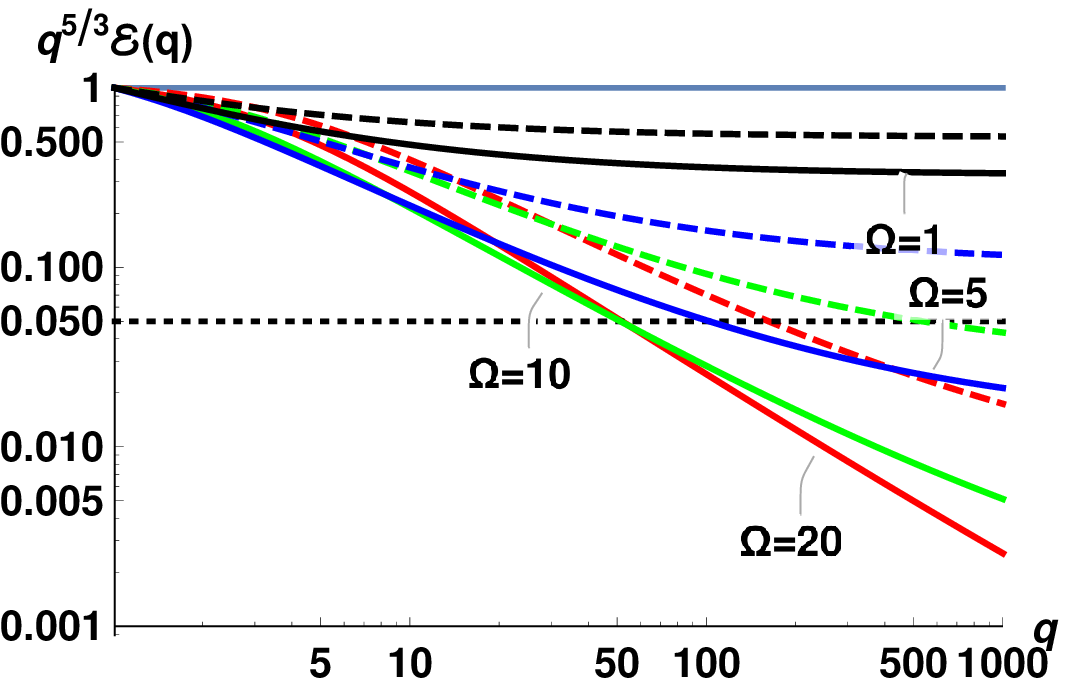}&
 			\includegraphics[scale=0.45 ]{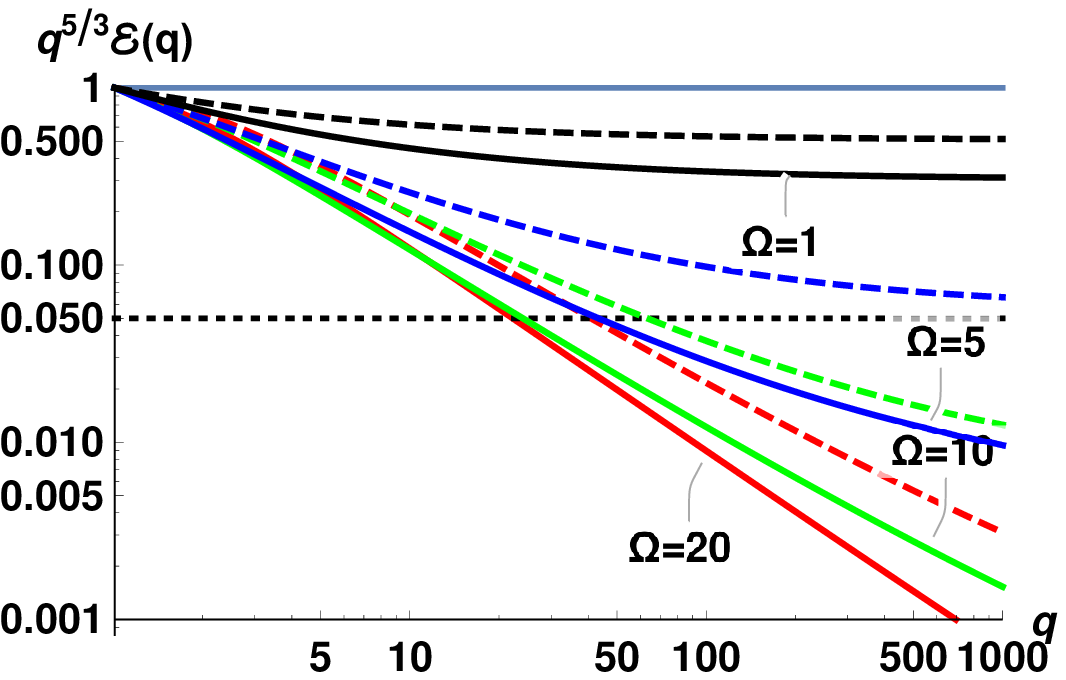}\\ 		 
 		$T=1.95\,$K, $ \ \~ U\sb {ns}=1.0 $& 	$T=1.95\,$K, \ $ \~ U\sb {ns}=4.0 $  &	$T=1.95\,$K,\  $ \~ U\sb {ns}=8.0 $ \\
 			\includegraphics[scale=0.45 ]{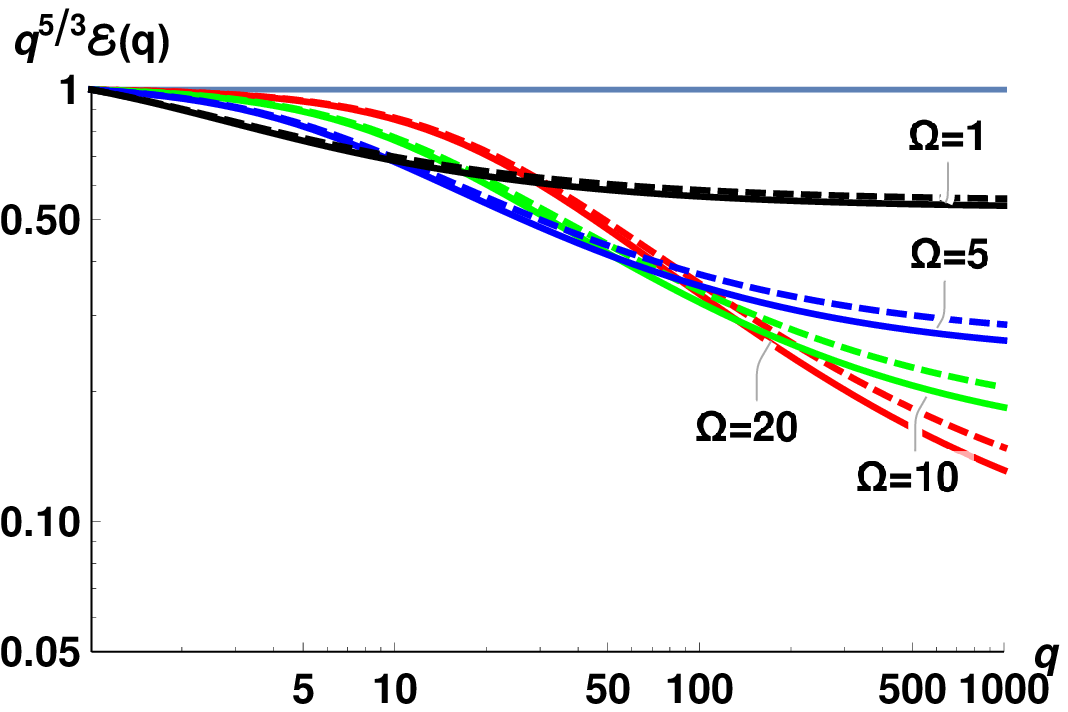}&
 			\includegraphics[scale=0.45 ]{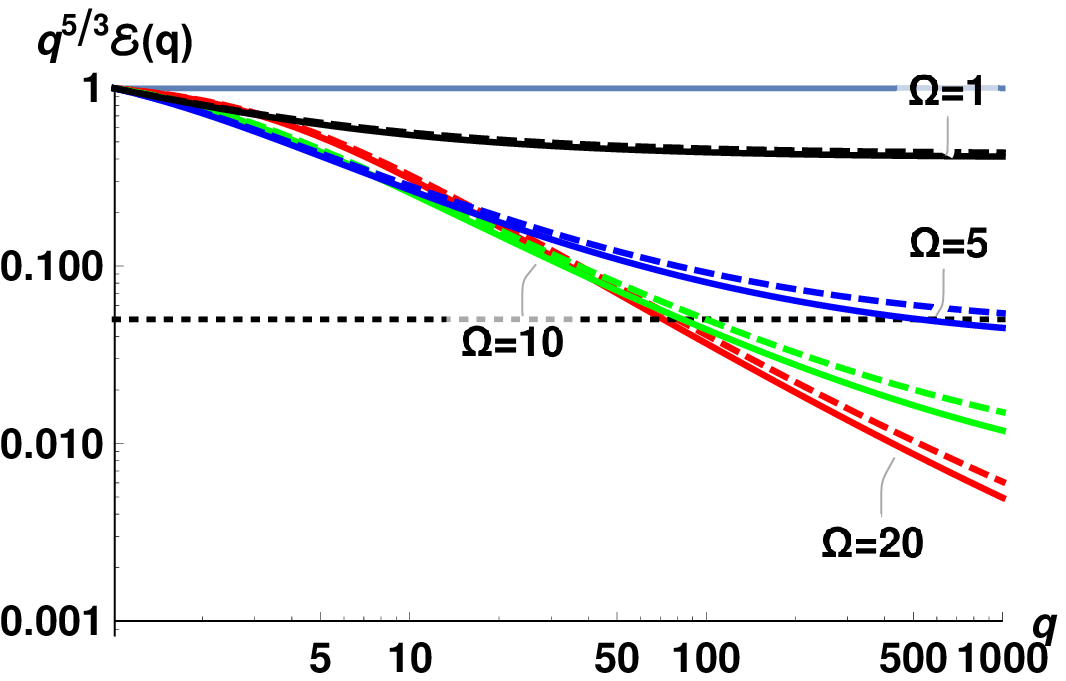}&
 			\includegraphics[scale=0.45 ]{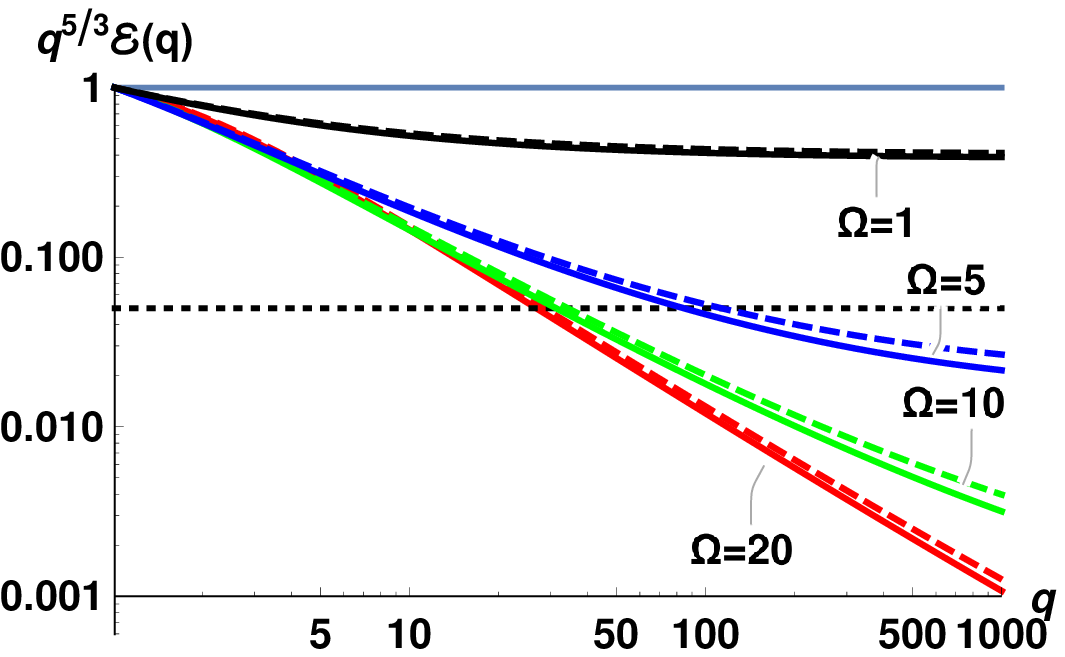}\\		 
 		$T=2.10\,$K, $ \ \~ U\sb {ns}=1.0 $& 	$T=T=2.10\,$K, \ $ \~ U\sb {ns}=4.0 $  &	$T=T=2.10\,$K,\  $ \~ U\sb {ns}=8.0 $ \\
 			\includegraphics[scale=0.45 ]{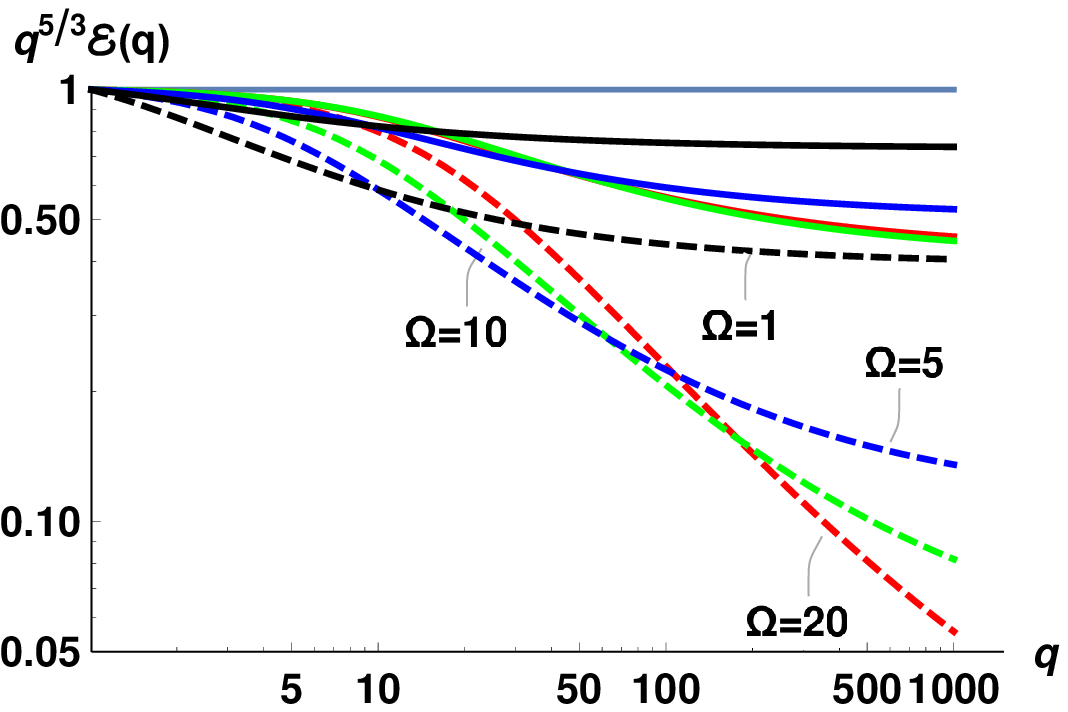}&
 			\includegraphics[scale=0.5 ]{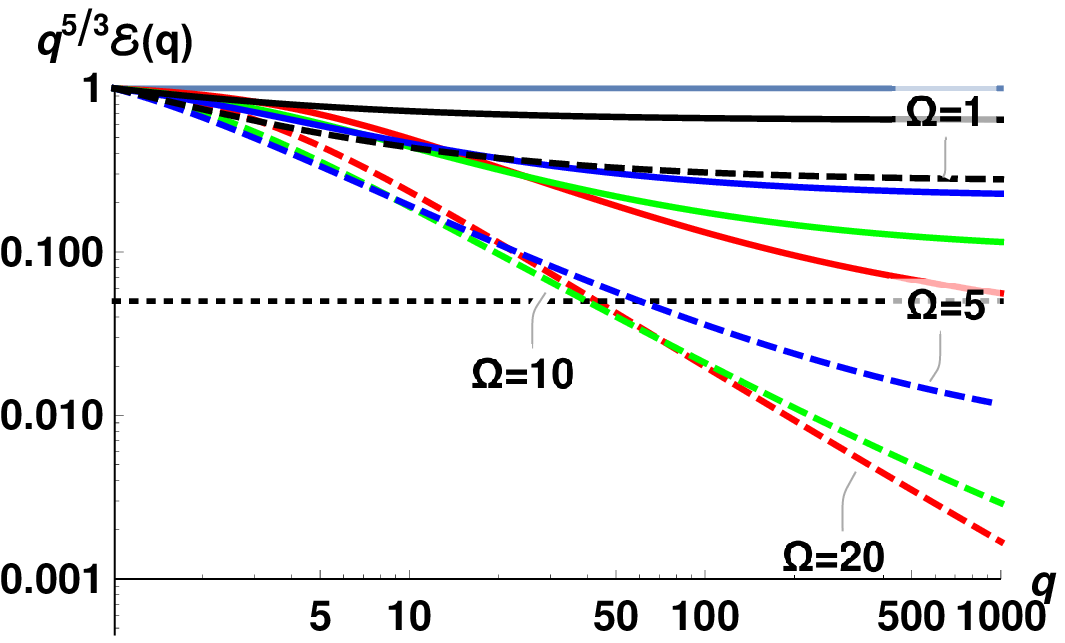}&
 		        \includegraphics[scale=0.46]{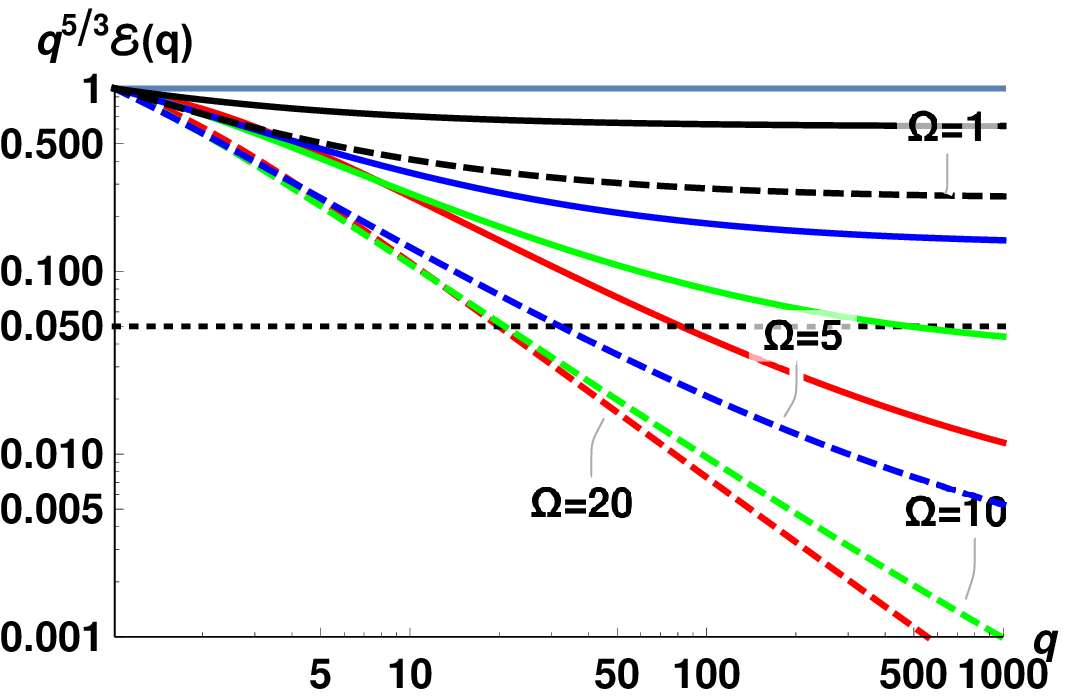}\\
 			
 		\end{tabular}
 		\caption{\label{ff:1} Color online. The compensated energy spectra 
 			$\C E_j(q)=q^{5/3} E_j(q)/E_j(q_0)$ vs $q=k/k_0$  for ${\rm Re}_j=\infty$ and different combinations of $T$ and $\tilde  U\sb{ns}$. 
 			The lines corresponding to different values of  $\~\Omega\sb{ns}$ are shown by different colors (from top to bottom): $\~\Omega=0$ (the horizontal gray lines),
 			$\~\Omega\sb{ns}=1.0$ (black lines),   $\~\Omega\sb{ns}=5.0$ (blue lines), $\~\Omega\sb{ns}=10.0$ (green lines) and  $\~\Omega\sb{ns}=20.0$ (the lowest red lines).
 			The  normal-fluid energy spectra are shown by solid lines, the superfluid
 			spectra -- by dashed  lines.
 			Note that in the left column  $q^{5/3}\C E_j(q)$ varies from 0.05 to
 			1.0, while in the middle and the right columns -- from 0.01 to 1.0. In these panels the
 			level 0.05 is shown by the horizontal dotted lines. The labels ''$\Omega$`` in the figures mark one of the lines of the corresponding color (solid or dashed) for further clarity. }
 	\end{figure*}
 	
 	\begin{figure*}
 		\begin{tabular}{ccc}
 			$\~U\sb{ns}=1$   &$\~U\sb{ns}=4$ K & 	$\~U\sb{ns}=8$   \\
 			\includegraphics[scale=0.47]{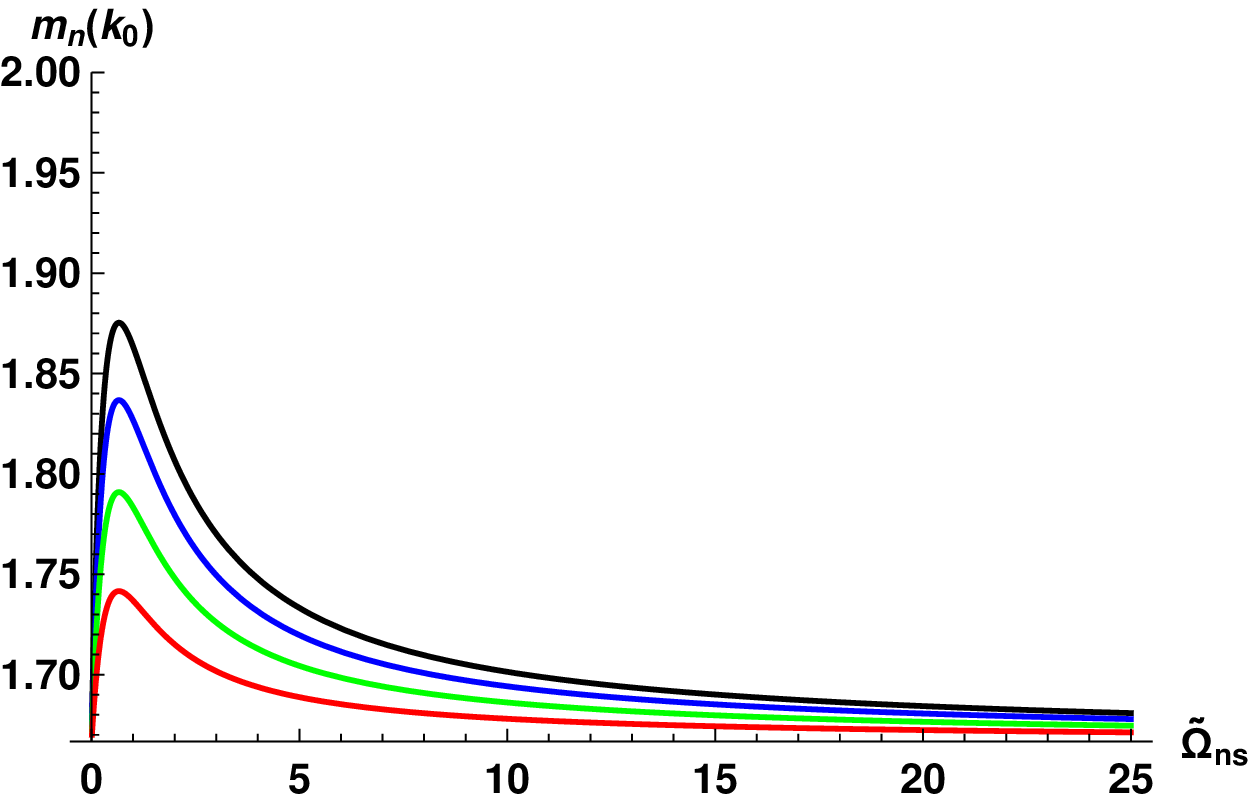}&
 			\includegraphics[scale=0.47]{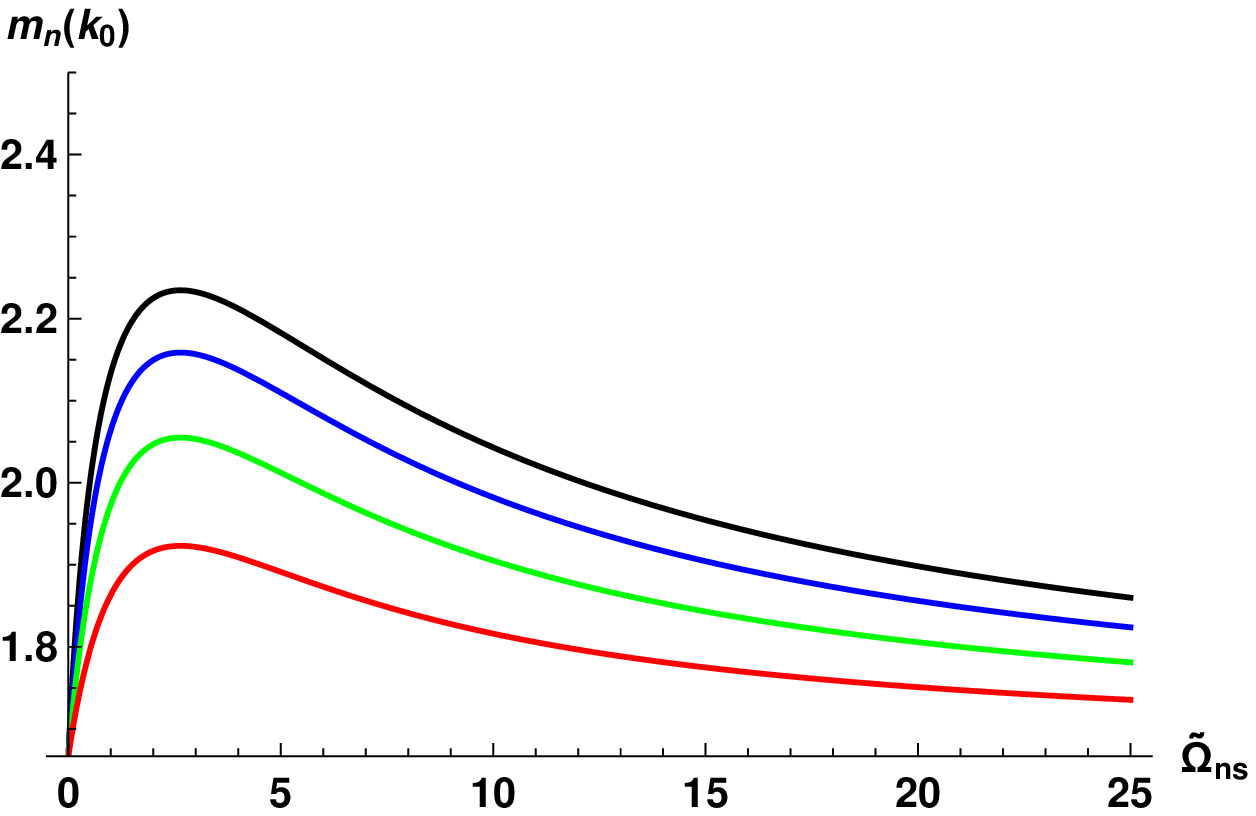} &
 			\includegraphics[scale=0.47]{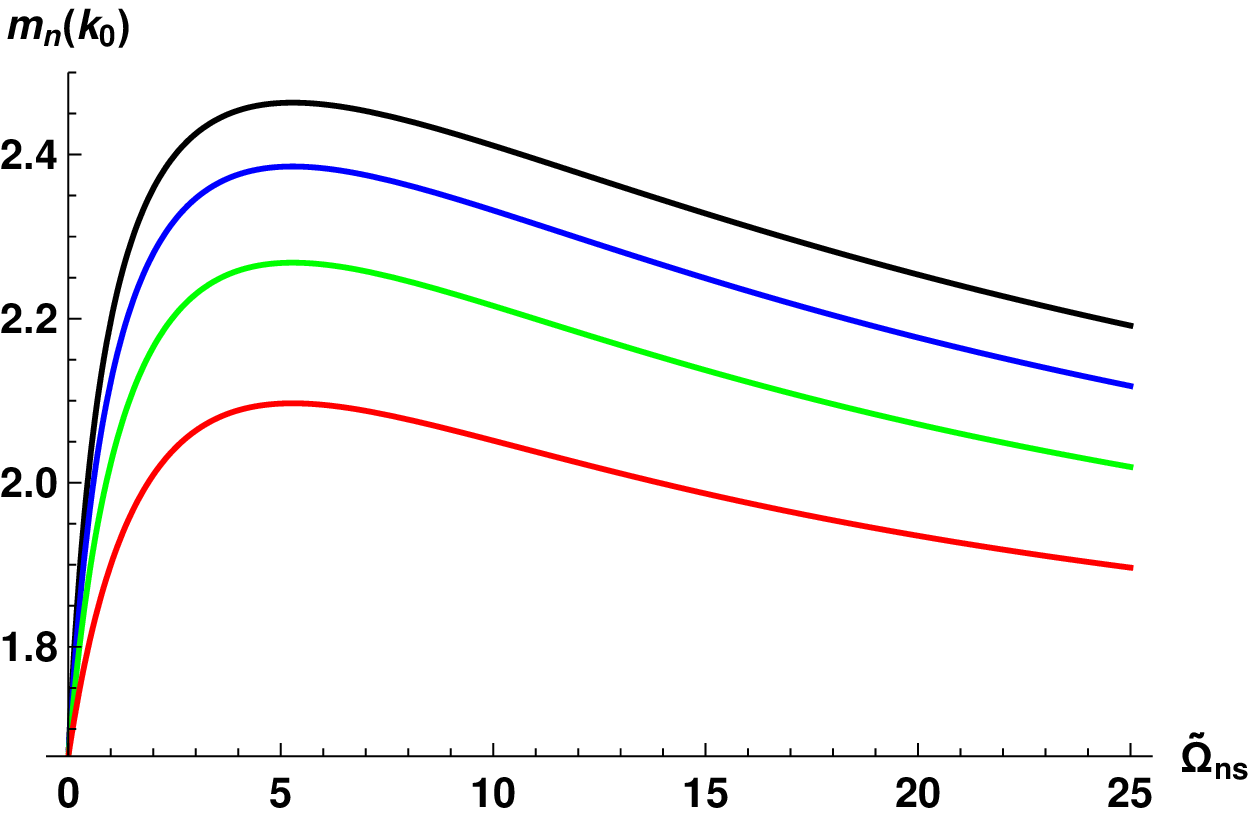}  \\
 		\end{tabular}
 		\caption{\label{ff:2} Color online.The outer-scale scaling exponents of the normal component $m\sb n(1)$ [\Eq{mj}] vs. $\~\Omega\sb {ns}$ for $\~U\sb{ns}=1$ (left panel),   $\~U\sb{ns}=4$ (middle panel) and  $\~U\sb{ns}=8$ (right panel). 
 			 The lines from top to bottom correspond to $T=1.65\,$K (black lines),  $T=1.85\,$K (blue lines),  $T=2.0\,$K (green lines), and $T=2.10\,$K (red lines).}  
 	\end{figure*} 
 
\begin{figure}
			\includegraphics[scale=0.6]{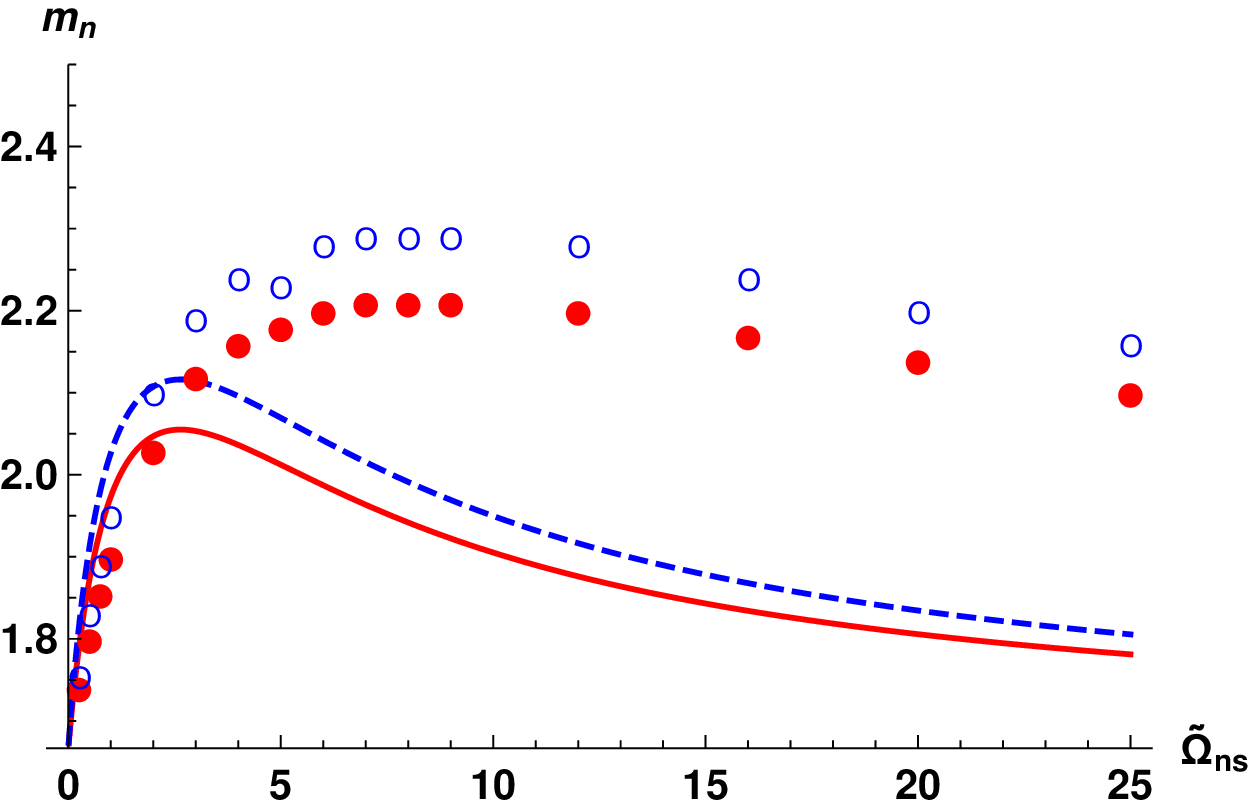} 
		 	\caption{\label{ff:2b} Color online. The $\~\Omega\sb {ns}$  dependence of the  scaling exponents  $m_j (1)$  and the mean exponent $\< m_j \>_{10}$. The normal component exponent $m\sb n(1)$ is marked as red solid line,  $\< m\sb n\>_{10}$ --by red full circles. The superfluid exponent  $m\sb s(1)$  is marked by blue dashed line, the mean exponent $\< m\sb s \>_{10}$ -- by empty blue circles. The exponents were calculated for $T=2.0\,$K and $\~U\sb {ns}=4$. 
		 	}
\end{figure}

\section{\label{s:ESCT} Energy spectra of counterflow turbulence}
 \subsection{\label{ss:QSB} Qualitative  analysis  of the energy spectra }
To qualitatively analyze the energy spectra, we first neglect in \Eqs{20} the influence of the viscous dissipation.
In  \Fig{ff:1} we show  the energy spectra, obtained by solving \Eq{20A} in a wide range of dimensionless parameters with  Re$\sb {s,n}\to \infty$. In all panels, the normal component spectra are shown by solid lines and the  superfluid component spectra -- by dashed lines.
 
 Each row represent the spectra at one of four temperatures, from top to bottom: $T=1.65,1.85, 1.95$ and $2.1$K.
 
 The three
 columns show  solutions  for three different values of the dimensionless counterflow velocity\,\eq{Uns}:  $\~U\sb{ns}=1$ (left), $ \~U\sb{ns}=4$ (middle) and $\~U\sb{ns}=8$ (right).
 
 Each of the
 panels contains the spectra for  5
 values of the dimensionless frequency $\~\Omega\sb{ns}$ from 0 (the horizontal gray lines) to $\tilde \Omega\sb{ns}=20$ (the lowest red lines), color-coded as described in the figure caption.

 Comparing spectra shown in  \Fig{ff:1}, we can make a set of observations: 
 
 	i) The larger  the counterflow velocity $\~U\sb{ns}$, the stronger is the 
 	suppression  of the energy spectra compared to the K41 prediction. This is an expected result: the 
 	normal-fluid and superfluid velocity fluctuations decorrelate faster  with increasing $\~U\sb{ns}$, leading to stronger energy dissipation  by mutual friction and as a result-- to a more prominent suppression of the energy spectra.
 	
	ii)  In the absence of viscous dissipation, the dissipation by mutual friction defines the suppression of the spectra.  The corresponding frequencies $\Omega_j$ [\Eq{20B}], are proportional to the other component's densities: $\Omega\sb s\propto \rho\sb n, \Omega\sb n \propto \rho\sb s$.
 	Therefore, at low temperatures (two upper rows), when $\rho \sb n \ll \rho \sb s$, the normal-fluid component spectra are suppressed stronger than the superfluid spectra, while at high $T$ the relation is reversed (the lowest row).
 	
 	 	iii) The competition between the velocity fluctuations coupling and the dissipation due to mutual friction leads to a complicated $q$-dependence of the spectra, described by  \Eq{Balance1A}: the rate of energy dissipation is proportional to $\Omega_j[D_j(q)-1]$. The larger is   $\Omega_j$ the stronger is the coupling, however, simultaneously $ D_j(q)\to 1$. Which factor wins, depends on the scale: at large $q$ the dissipation wins and the spectra suppression is directly proportional to $\~\Omega_j$. On the other hand, at small $q$, the spectra for larger  $\~\Omega_j$ are less suppressed, especially at weak counterflow velocity. 
 
 Note that the applicability range of HVBK [\Eqs{NSE}] is limited by  $k<\pi/\ell=\pi \sqrt{\C L}\simeq \pi \sqrt{2\Omega\sb{ns}/\kappa}$. However  our analysis is performed for given values of $\~\Omega\sb{ns}=\Omega\sb{ns}/k_0 u\sb T$ with $k_0=1$ and an arbitrary value of $u\Sb T$. Therefore, there is no formal restriction on the range of $k$  in \Figs{ff:1}, (as well as in \Figs{ff:2}, \ref{ff:2b}, and  \Fig{ff:3}),  which was chosen as to expose all important features of the energy spectra.
 
\begin{figure*}
	\begin{tabular}{ccc}
		$T=1.65\,$K   &	$T=1.95\,$K & 	$T=2.10\,$K  \\
		\includegraphics[scale=0.4]{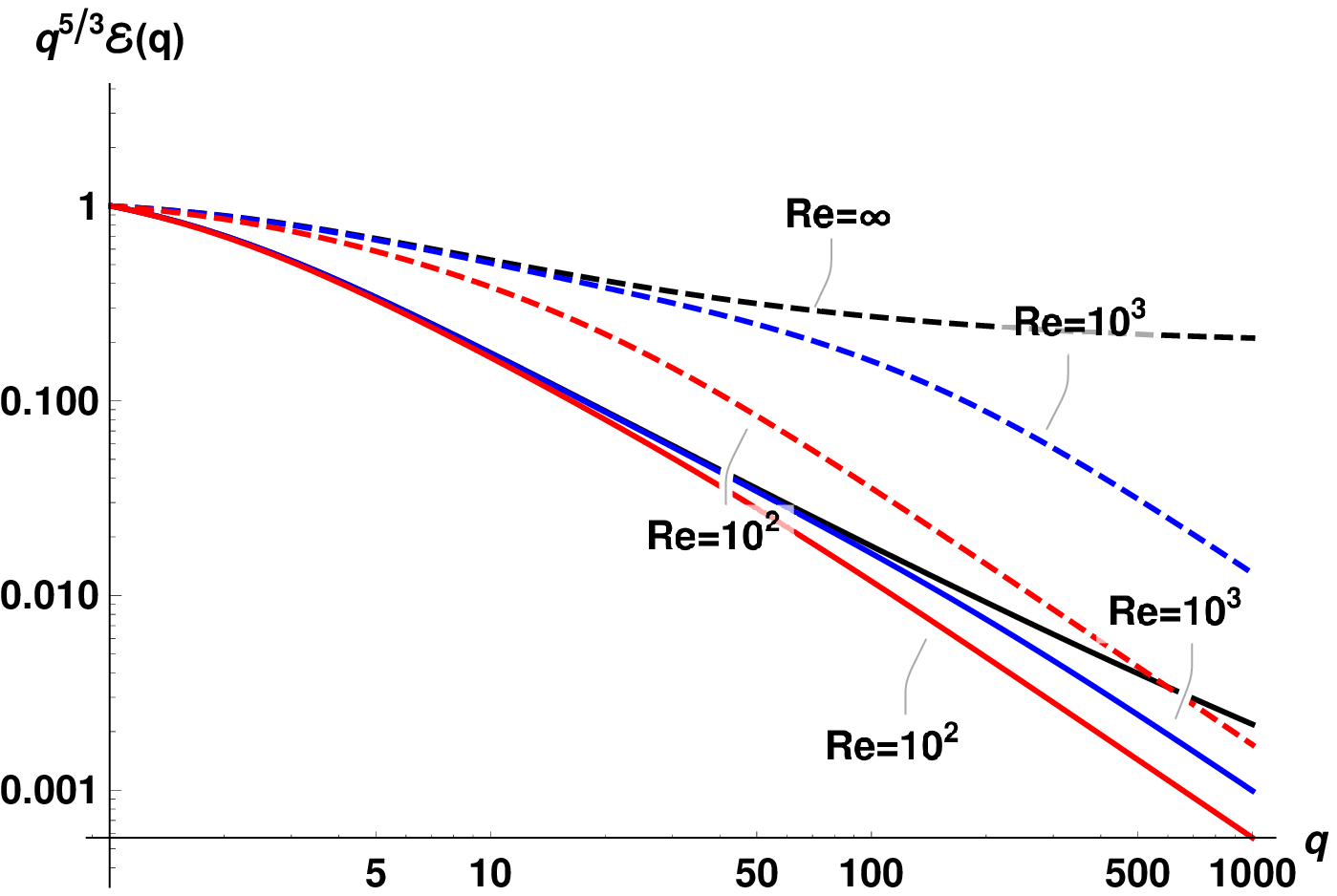}&
		\includegraphics[scale=0.4]{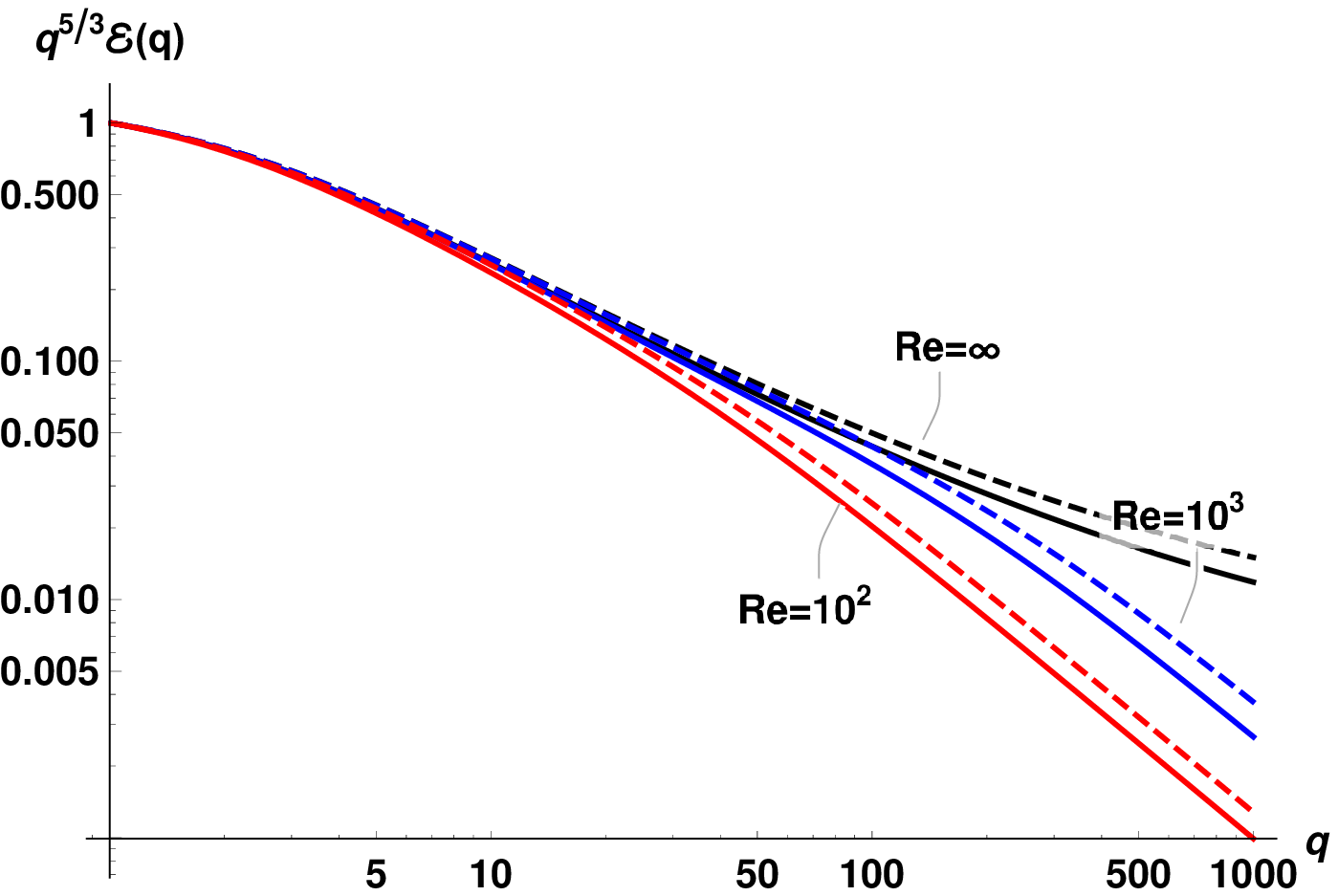} &
		\includegraphics[scale=0.4]{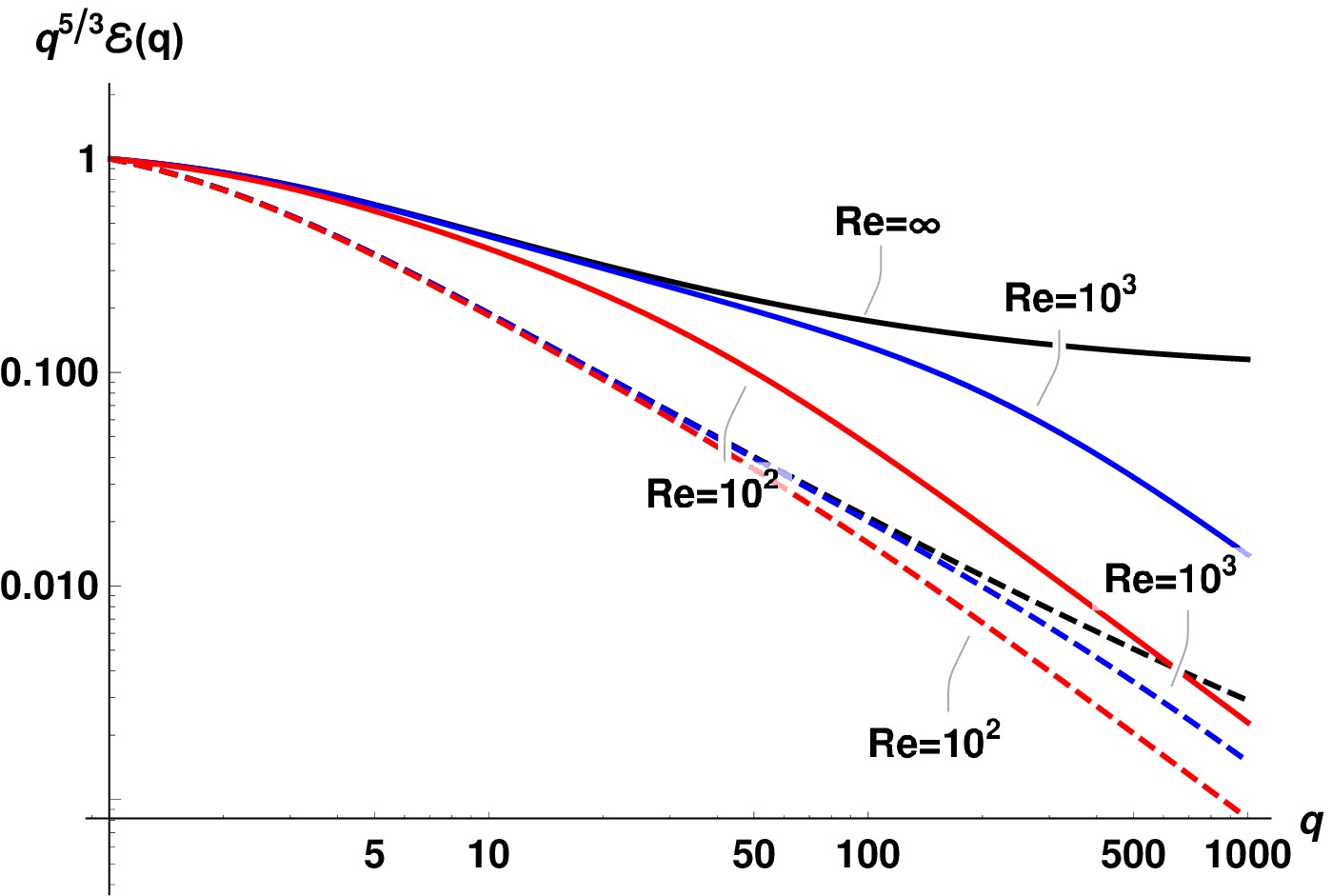}  \\
	\end{tabular}
	\caption{\label{ff:3} Color online.  The viscous corrections to the
		energy spectra at different temperatures: $T=1.65\,$K (left panel), $T=1.95\,$K (middle panel) and  $T=1.21\,$K (right panel). The norma-fluidl spectra are shown by solid lines, the superfluid spectra -- by dashed lines.  The spectra are shown for  Re$\sb n=\infty$ (black lines),  Re$\sb n=10^3$ (blue lines) and $ 10^2$ (red  lines).  All spectra were calculated with $\~U\sb{ns}=4$ and $\~\Omega\sb{ns}=10$. }
\end{figure*} 

  \begin{figure}
 	\includegraphics[scale=0.6]{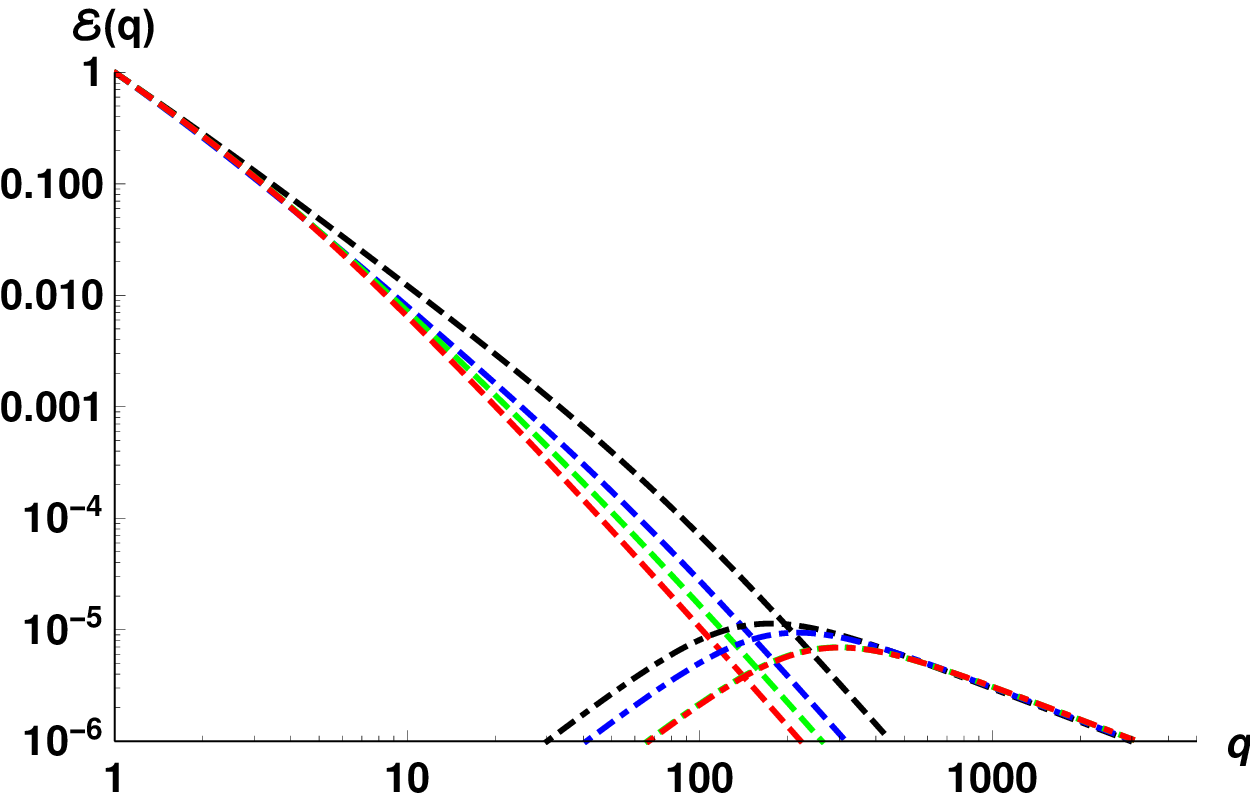}
 	\caption{\label{ff:6} Color online. The (uncompensated) energy spectra of counterflow turbulence, $\C E\sb s(q)=
 		E(q)/E(q_0)$ vs $q=k/k_0$   (dashed lines)  and the sketch of the quantum peak
 		(dot-dashed lines) for $T=1.65\,$K,  $Q=300\,$mW/cm$^2$ (upper black lines)
 		$T=1.85\,$K, $Q=497\,$mW/cm$^2$ (blue lines) $T=2.00\,$K  $Q=586\,$mW/cm$^2$
 		(green lines) and $T=2.10\,$K  $Q=350\,$mW/cm$^2$ (lowest red lines). Note that here the $q$-range, where $\C E\sb s(q)$ are  valid, is limited by $q\lesssim 200$.  }
 \end{figure}

 \subsection{\label{ss:m} Outer-scale and mean scaling exponents of the energy spectra}
 To characterize  in a compact form  the energy spectra dependence on the flow parameters, $T$, $\~U\sb{ns}$ and $\~\Omega\sb{ns}$, we consider first a scaling exponent $m(q)$, \Eq{LP17b},  in the beginning of the scaling interval $k=k_0, q=1$. Using \Eqs{psiA} and \eqref{20} we get for outer-scale exponent $m_j(q=1)$:
 \begin{eqnarray}\label{mj}
  m_j(1)=\frac53+\frac43\, \frac{\~\Omega_j\big [  1-q_\times \arctan(1/q_\times)\big ]}{4 C+\~\Omega_j\big [  1-q_\times \arctan(1/q_\times)\big ]  }\,.~~~~~
 \end{eqnarray}
 
 The   $\~\Omega\sb{ns}$-dependencies of the normal-fluid outer-scale exponent  $m_n(1)$  for different $\~U\sb{ns}$ and $T$ are shown in \Fig{ff:2}. As expected,  for $\~\Omega\sb{ns}=0$ (no mutual friction damping) $m\sb n (1)=\frac53$, the classical K41 value. In the limit $\~\Omega\sb{ns}\to \infty$, $\displaystyle \lim_{q_\times\to \infty}[q_\times \arctan(1/q_\times)]\propto 1/q_\times^2$
 and $m\sb n(1)$ again tends to its K41 value $\frac53$. In this limit, the normal and superfluid components are fully coupled and there is no energy dissipation by  mutual friction.  The resulting $\~\Omega\sb{ns}$-dependence of $m_n(q)$ is non-monotonic with a maximum around $\~\Omega\sb{ns}\sim 1$.  As we saw before, the  $m_n(1)$  is largest (i.e. the strongest suppression of normal-fluid energy spectra) } at lowest $T=1.65\,$K (upper black lines), the smallest -- at high $T=2.1\,$K (the lowest red lines). 
 
 Conversely,  the superfluid exponents  $m_s(1)$ (not shown) reflect  strong  suppression [larger $m_s(1)$]  at high temperatures, while at low temperatures $m_s(1)$ are smaller. 
 
To characterize the degree of deviation from the scale invariance, we introduce the mean scaling exponent over some $q$-interval from  $q=1$ [with $\C E_j(1)=1$] to  a given value of $q$ : 
 \begin{equation}\label{meanM}
 \< m_j\>_q= - \ln \C E_j(q) \big / \ln q\ .
 \end{equation}
 The value $\< m_j\>_q$ should be $q$-independent and equal to the outer-scale exponent $m_j(1)=[d \ln \C E_j(q)/ d \ln q]_{q=1}  $ for the scale-invariant spectra and vary otherwise.  In \Fig{ff:2b} we compare $m_j(1)$ and $\< m_j\>_{10}$ for $T=2.0$K and $\~U\sb{ns}=4$. The  outer-scale scaling exponent $m\sb n(1)$ is shown by solid line,  $m\sb s(1)$ -- by dashed line. The values of  $\< m\sb n\>_{10}$  and $\< m\sb s\>_{10}$, calculated for a set of $\~\Omega\sb{ns}$,  are shown by full and empty circles, respectively. Clearly, the spectra coincide (at least within the first decade) for $\~\Omega\sb{ns} \lesssim 3$, while for stronger  $\~\Omega\sb{ns}$ they vary significantly: the actual spectra are more suppressed than is suggested by $m_j(1)$ and $\< m_j\>_{10}> m_j(1)$ for $\~\Omega\sb{ns}>3$ at this temperature.  The non-monotonic behavior is evident  in all curves, although the maximum in the mean  exponents $\< m_j\>_{10}$  is broader and less prominent.

 \subsection{\label{ss:Re}  Viscous damping of the energy spectra }
Having in mind the influence of $\~\Omega\sb{ns}$ and $\~U\sb{ns}$ on the energy spectra, we now add the viscous dissipation to the picture and plot in \Fig{ff:3} the compensated spectra  for $T=1.65,  \ 1.95\,$ and $T=2.10\,$K, using $\~U\sb {ns}=4$ and $\~\Omega\sb{ns}=10$. The spectra for Re$\sb n \to \infty$ are shown by black lines, those for Re$\sb n=10^3$-- by blue lines, while red lines denote the spectra for Re$\sb n=10^2$. The corresponding Re$\sb s$ were calculated using the ratio of viscosities (cf. Table \ref{t:1}).
 
As expected, the reduction of Re$  _j$ leads to progressive damping of the energy spectra.  This effect is mostly concentrated at large $q$, while  at small $q$  the energy dissipation by mutual friction  dominates. The viscous damping is most prominent for the denser component (superfluid at $T=1.65$K and normal fluid at $T=2.1$K). At $T=1.95\,$K, the densities of the components are close and the spectra are similar for all Re numbers.

 \subsection{\label{ss:QP} Quantum peak in superfluid energy spectra}
 As has long been known \cite{Vinen3}
 and understood \cite{Schwarz} that besides large scale turbulence, discussed above, the intense counterflow generates the superfluid turbulence (the vortex tangle).
 The corresponding energy spectra  peak at the intervortex scale
 $\ell=1/ \sqrt{\C L}$.  This kind of superfluid turbulence has no classic analog
 and is traditionally called Vinen's or ultra-quantum turbulence.
 
The total energy density of quantum turbulence (per unit mass of superfluid component) $E\Sb Q$ may be reasonably estimated within the Local Induction Approximation\cite{Schwarz} as
 \begin{subequations}\label{EQ}
 	\begin{equation}\label{EQa}
 	E\Sb Q = \frac{\kappa^2 \C L \Lambda}{2\pi}\,, \quad \Lambda\approx \ln
 	(\ell/a_0)\ .
 	\end{equation}
 	Here $a_0$ is the vortex core radius ($\sim10^{-8}\,$cm in $^4$He). For the 
 	typical value $\C L\simeq 10^5$, $\Lambda \approx 12.6$ and very weakly depends on
 	$\C L$. Therefore for our purposes we can estimate
 	\begin{equation}\label{EQb}
 	E\Sb Q \simeq 2 \,\kappa^2 \C L \ .
 	\end{equation}
 \end{subequations}
 Using   experimental values of  $\C L$,  discussed below,  we found $E\Sb Q$ and compared them with $E\sb {cl}$, calculated for the experimental conditions.
 It is interesting to realize that $E\sb {cl}\sim E\Sb Q$.  For example, at $T=1.65\,$K the ratio $E\Sb
 Q/E\sb{cl}$ varies between 1.2 and 1.8, for $T=2.00\,$K  $E\Sb
 Q/E\sb{cl}\simeq 5$. This  fact may be rationalized by  simple
 models of turbulent channel flow (cf. e.g \Ref{Pope}) and by dimensional reasoning.
 Indeed, for the classical channel flow, the  dimensional reasoning (and simple models -- up
 to logarithmic corrections) give $\Delta u\sb n \simeq U\sb {ns}$ as supported by the experiment [column \# 8 of Tab.\,\ref{t:2}].
 Thus $E\sb{cl}\simeq U\sb {ns}^2$. 
 
 For the quantum energy of superfluid turbulence, 
 \Eqs{EQb} gives $E\Sb Q\simeq 2 \kappa^2 \C L$, while $\C L  = (\gamma
 U\sb{ns})^2$ with $\gamma \simeq 1/\kappa$. Therefore, also for the quantum energy  $E\Sb Q\simeq U\sb
 {ns}^2$.
 
 Our knowledge of quantum peak $k$-dependence, $E\Sb Q(k)$, is quite
limited. Dimensional reasoning, supported by the numerical simulations\cite{Baggaley2012},
 predicts maximum of $E\Sb Q(k)$ at the inverse intervortex distance $k_*\simeq 2\pi
 \sqrt {\C L}$. For $k_* \ll  k < 2\pi /a_0$, $E\Sb Q(k)$ is
 dictated by the velocity field near the vortex line: $v(r)\propto \kappa/r$, where $r$
 is the distance to the vortex line. Modeling quantum vortex tangle as a set of randomly oriented vortex lines with the vortex line density $\C L$ and averaging  over line orientations, we get 
   an asymptotic behavior  $E\Sb
 Q(k)\simeq \kappa^2 \C L /k$ for $2\pi/a_0\gg k\gg  k_*$.  The same answer follows from dimensional reasoning based on a natural assumption that    $E\Sb
 Q(k)\propto \C L$.
 
 For  $k\ll  k_*$ we do not expect
 inverse energy cascade in 3D turbulence. Therefore, following \Ref{BLPSV-2016}, we
 assume here local thermodynamic equilibrium spectra with equipartition of energy
 between degrees of freedom:  $E\Sb Q(k)\propto k^2$. Simple analytic formula
 that reflects all these properties has a form:
 \begin{subequations}\label{guess}
 	\begin{equation}\label{guessA}
 	E\Sb q(k)= \frac {E\Sb Q}{\Lambda}\, \frac {k^2}{k_*^3+k^3}\,, \  k_*=
 	\frac{2\pi}{\ell}\ .
 	\end{equation}
 	Here $E\Sb Q$ is the total energy of quantum peak,
 	\begin{equation}\label{guessB}
 	E \Sb Q=\int_0^{2\pi/a_0}  E\Sb Q(k)\, dk\ .
 	\end{equation}
 \end{subequations}
 
 Taking the values of $ E \Sb Q$ and $\ell=1/\sqrt {\C L}$ from the
Tab. II, we plot in Fig. \ref{ff:6}
 the  energy spectra, corresponding to the quantum-peak\,\eqref{guess} for four temperatures (dot-dashed lines). We also
 show  by dashed lines the quasi-classical superfluid energy spectra $E\sb s(k)$. We see that the quasi-classical and quantum parts of the superfluid energy spectra are well separated in the
 $k$-space, as was suggested in \Refs{{BLPSV-2016},Vinen} for the explanation of the
 vortex line density decay $\C L(t)$ after switching off the counterflow. 
 
 What is important for us now is that the distinct separation of the quasi-classical
 and quantum contributions to the  superfluid energy spectra allows us to neglect
 the direct effect of the quantum peak on the  behavior of the normal fluid and
 superfluid quasi-classical turbulence. The only role played by the quantum peak
 in our theory is to give an independent and leading contribution to the vortex-line density that
 determines the mutual friction.

 \subsection{\label{s:details} Energy spectra in the conditions of the Tallahassee  experiments}
 
 Now we are ready to analyze the energy spectra for conditions, close to realized in the $^4$He counterflow visualization experiment\cite{WG-17a}.
 
 The experiments\cite{WG-17a,WG-17b} in the turbulent counterflow of superfluid
 $^4$He   were conducted for a range of temperatures and heat fluxes. A number of
 important properties of the flow, required for comparison between theory and
 experiment are listed  in Table~\ref{t:2}.
 
 The normal velocity fluctuations were deduced by the visualization of the
 molecular tracers\,\cite{WG-2015,WG-17a}. The ratio of this  turbulence
 intensity to the mean normal velocity is almost independent of the values of
 the heat flux for a given temperature\,\cite{WG-17a}.
 The vortex line density $\C L$ was measured by the second sound attenuation. 
 \begin{table*}
 	\begin{tabular}{| c |  c |  c |  c | c | c|  c | c|   c  | c  |c | c|  c| c|}
 		\hline\hline
 		\# & 1        & 2              & 3         & 4                                & 5                     & 6                          & 7                           & 8                     & 9                               & 10           & 11              & 12               & 13                \\ \hline
 		& ~~$T,$~~ & $Q, $          & $U\sb n,$ & $ u\Sb{T }$ & ${\cal L},$           & $ {\rm Re}\sb n=$          & ${\rm Re}\sb s=$            &  $\~ U\sb{ns}=$   & $\~ \Omega\sb{ns}=$        & $E\sb{cl},$  &  $E\Sb Q,$  & {$n\sb{exp}+1$}     &  $\<m\sb n\>\sb{10 }$ \\
 		\# & ~~K~~    & $\rm mW/cm^2 $ & cm/s      &   cm/s                              & cm$^{-2}$             & $  u \Sb T/(k_0 \nu\sb n)$ & $ \ u \Sb T/(k_0 \nu\sb s)$ & $\ U\sb{ns}/ u\Sb{T}$ & $\ \Omega\sb{ns}/(k_0 u\Sb{T})$ & cm$^2$/s$^2$ & cm$^2$/s$^2$    &                  &                   \\ \hline\hline 	
 		1  &          & 150            & 1.87      &0.5                            & 8.63 $\times 10^4$    & 37.89                      & 76.78                       & 4.64                  & 3.47                            & 0.12         & 0.17            & 1.89$\pm0.03$    & 2.48              \\
 		2  & 1.65     & 200            & 2.23      &0.61                              & 16.2 $\times 10^4$    & 46.23                      & 93.67                       & 4.52                  & 3.35                            & 0.18         & 0.32            & 2.14$\pm0.03$    & 2.47              \\
 		3  & ~        & 300            & 3.27      & 1.12                               & 38.2  $\times  10^4$  & 84.88                      & 171.99                      & 3.61                  & 6.87                            & 0.62         & 0.73            & 2.18$\pm0.04$    & 2.43              \\ \hline
 		4  &          & 200            & 1.18      & 0.38                            & 8.11 $\times 10^4$    & 53.22                      & 50.21                       & 4.87                  & 3.72                            & 0.07         & 0.16            & 1.88$\pm0.04$    & 2.41              \\
 		5  & 1.85     & 300            & 1.78      & 0.67                               & 19.8 $\times 10^4$    & 94.21                      & 88.52                       & 4.17                  & 5.14                            & 0.22         & 0.39            & 2.23$\pm0.02$    & 2.38              \\
 		6  & ~        & 497            & 3.03      & 1.17                               & 58.5 $\times 10^4$    & 165.52                     & 154.59                      & 4.07                  & 8.71                            & 0.68         & 1.09            & 2.35$\pm0.03$    & 2.37              \\ \hline
 		7  &          & 233            & 0.86      & 0.44                            & 14.1$\times 10^4$     & 84.65                      & 51.01                       & 4.37                  & 5.66                            & 0.096        & 0.28            & 2.3$\pm0.02$     & 2.34              \\
 		8  & 2.0      & 386            & 1.34      & 0.68                                & 47.3$\times 10^4$     & 130.82                     & 78.84                       & 4.41                  & 12.29                           & 0.23         & 0.89            & 2.31$\pm0.03$    & 2.32              \\
 		9  & ~        & 586            & 2.09      & 1.16                              & 112 $\times 10^4$     & 223.17                     & 134.49                      & 4.03                  & 17.05                           & 0.67         & 2.04            & 2.36$\pm0.02$    & 2.23              \\ \hline
 		10 & {2.1 }   & {200}          & {0.57}    & 0.51                          & {37.3$\times 10^4$ }  & 106.93                     & 41.82                       & 4.31                  & 16.62                           & { 0.13}      & 0.71            & 2.09$\pm0.02$    & 2.08            \\
 		11 & ~        &  \,350        &   0.99     & 1.01                               &  114$\times 10^4 $    & 211.76                     & 82.82                       & 3.79                  & 25.65                           & {0.51}       & 2.07            & 2.11$\pm0.04$    &   1.96             \\ \hline\hline
 	\end{tabular}
 	\caption{\label{t:2} Columns \#\# 1--5: the experimental  parameters of the flow\cite{WG-17a}.  Columns \#\# 6--9:  the parameters of the model.  In columns \#10 and \#11: $E\sb{cl}=( u\sb T)^2/2$ and
 		$E\sb Q\approx 2   \kappa^2\, \C L$. Columns \#12 and \#13: the experimental values of the apparent scaling exponent $n\sb{exp}+1$ , and the theoretical normal fluid mean exponents over first decade $\<m\sb n\>\sb{10}$.}
 \end{table*}

 \subsubsection{\label{ss:exp-par} Measured and estimated parameters of the experiments }

The experiments\cite{WG-17a} were performed at four temperatures $T=~1.65, 1.85,2.0$ and $2.10\,$K using different values of the heat flux  $Q$, ranging from 150 to $\sim 600\,$mW/cm$^2$. The measured values of the resulting mean normal fluid velocity $U\sb n$, the normal-fluid rms turbulent velocity fluctuation $u\Sb T$ and the vortex line densities $\C L$ (columns \#3-\#5, Tab.\,\ref{t:2}) are reproduced according to Table I, \Ref{WG-17a}.
 
 Using these data and parameters of superfluid $^4$He for relevant temperatures, Tab.\,\ref{t:1}, we computed
 the ``turbulent" Reynolds numbers Re$_j$
 and listed them in columns \#6 and \#7 of Tab.\,\ref{t:2}. 
We used a 
 simplified assumption that  at large, energy containing scales, the rms turbulent velocity fluctuations $u_j$ of the normal and superfluid components  are close due their coupling by mutual friction: $u\sb s \approx u\sb n= u\Sb T$. As an estimate of the outer scale of turbulence we  take  $\Delta=0.225\,$cm which is a mean upper limit of the approximate scaling behavior of  $S_2(r)$, measured in \Ref{WG-17a}. Note that  the values of  Re$_j$ in these experiments are quite low,  with  Re$\sb n$ ranging  from Re$\sb n\simeq 38$ (line \#1) to  Re$\sb n\simeq 223$ (line \#9).
 
 The counterflow velocity $U\sb {ns}=U\sb n - U \sb s$ was found from the measured mean normal-fluid velocity $U\sb n$ and the condition of zero mass flux.
 
 Its resulting dimensionless  values  $\~ U\sb {ns}=U\sb {ns}/ u\Sb T$ are given in the column \#8.  
 
 The mutual-friction frequencies $\Omega\sb{ns}$ were calculated from \Eq{Ens1D}, using measured values of the VLD $\C L$  and $^4$He parameters. The dimensionless values $\~\Omega\sb {ns}$ are listed in the column  \# 9. They are ranging  from $\approx 3.4$ for $T=1.65\,$K to $\approx 26$ for $T=2.10\,$K.  We used the estimate $k_0\approx 2\pi / \Delta= 28\,$ cm$^{-1}$.
 
 \subsubsection{\label{ss:app} The scaling behavior of the energy spectra and the 2$\sp {nd}$ order structure function }
\begin{figure*}
 	\begin{tabular}{cc}
 	(a) $T=1.65\,$K & 	(b) $T=1.85\,$K   \\
 		\includegraphics[scale=0.6]{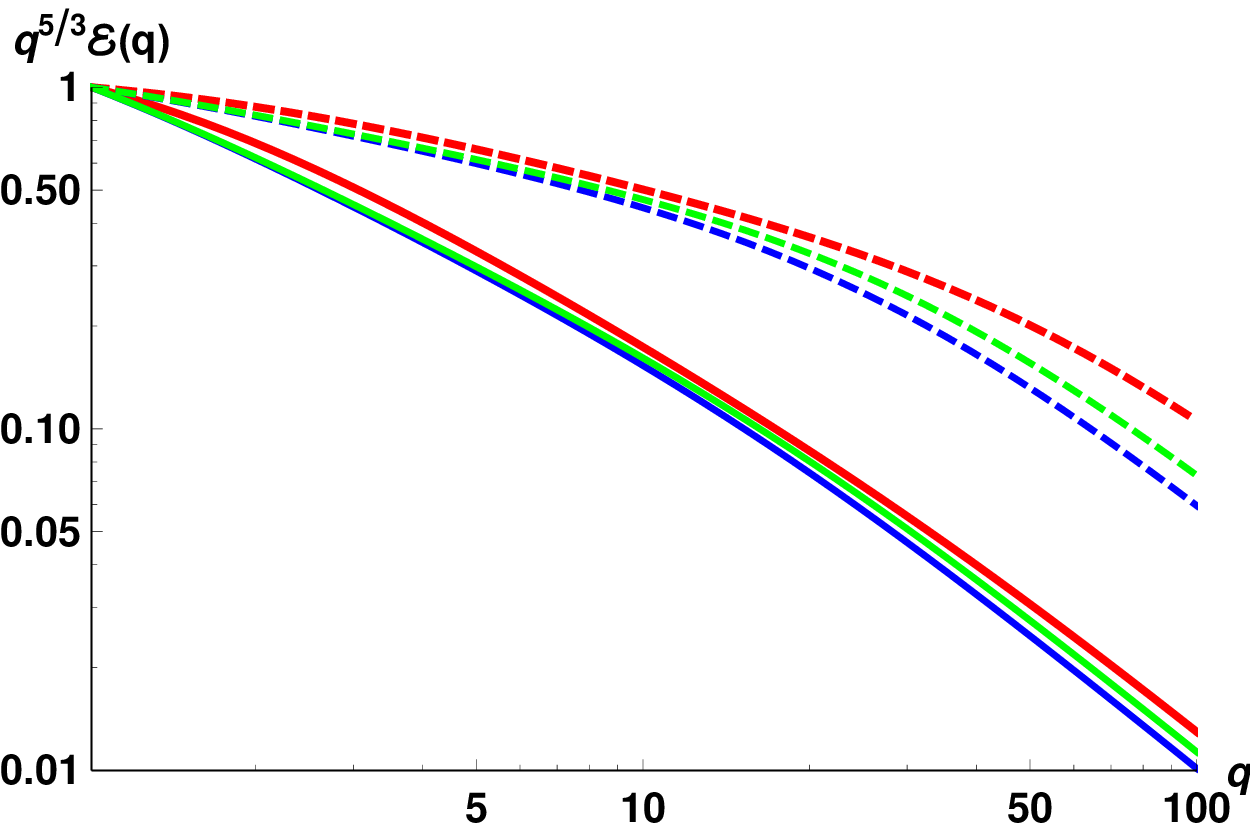}&
 		\includegraphics[scale=0.6]{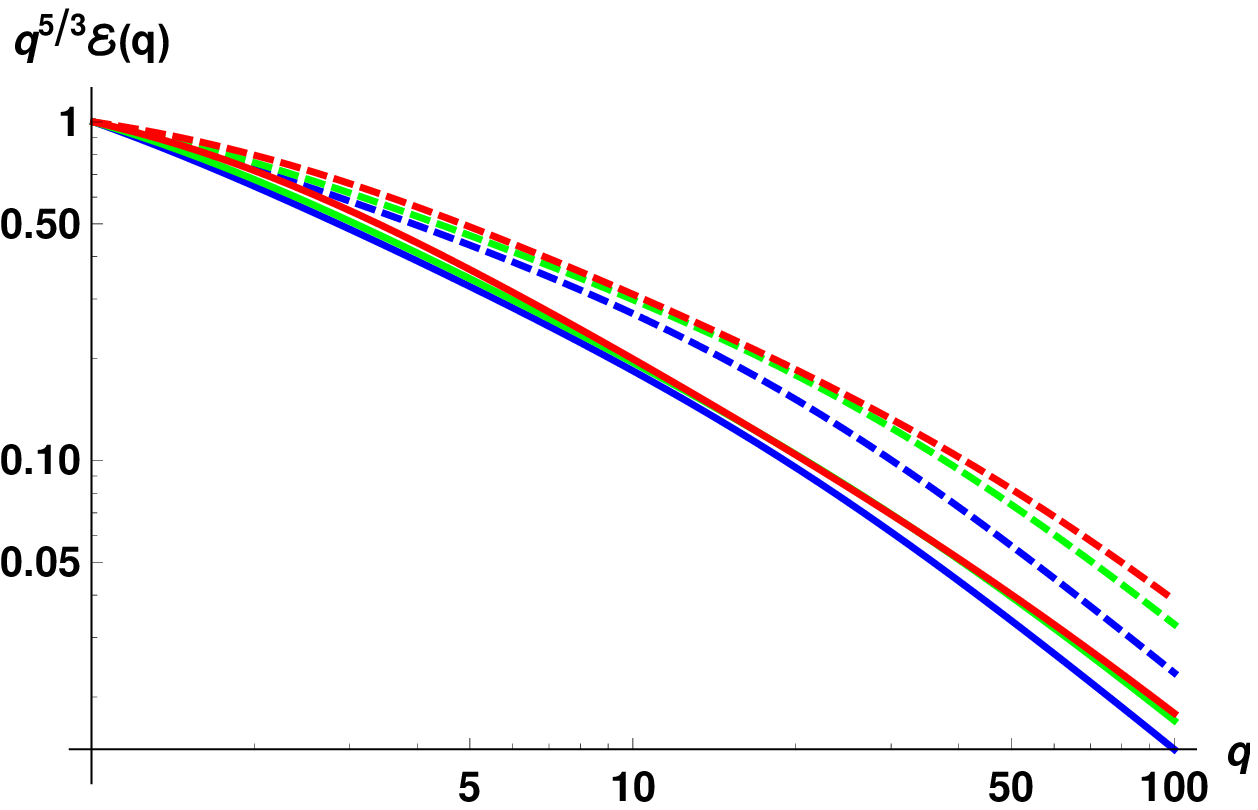} \\ & \\
 	(c)	$T=2.0\,$K & 	(d) $T=2.1\,$K   \\
 		\includegraphics[scale=0.6]{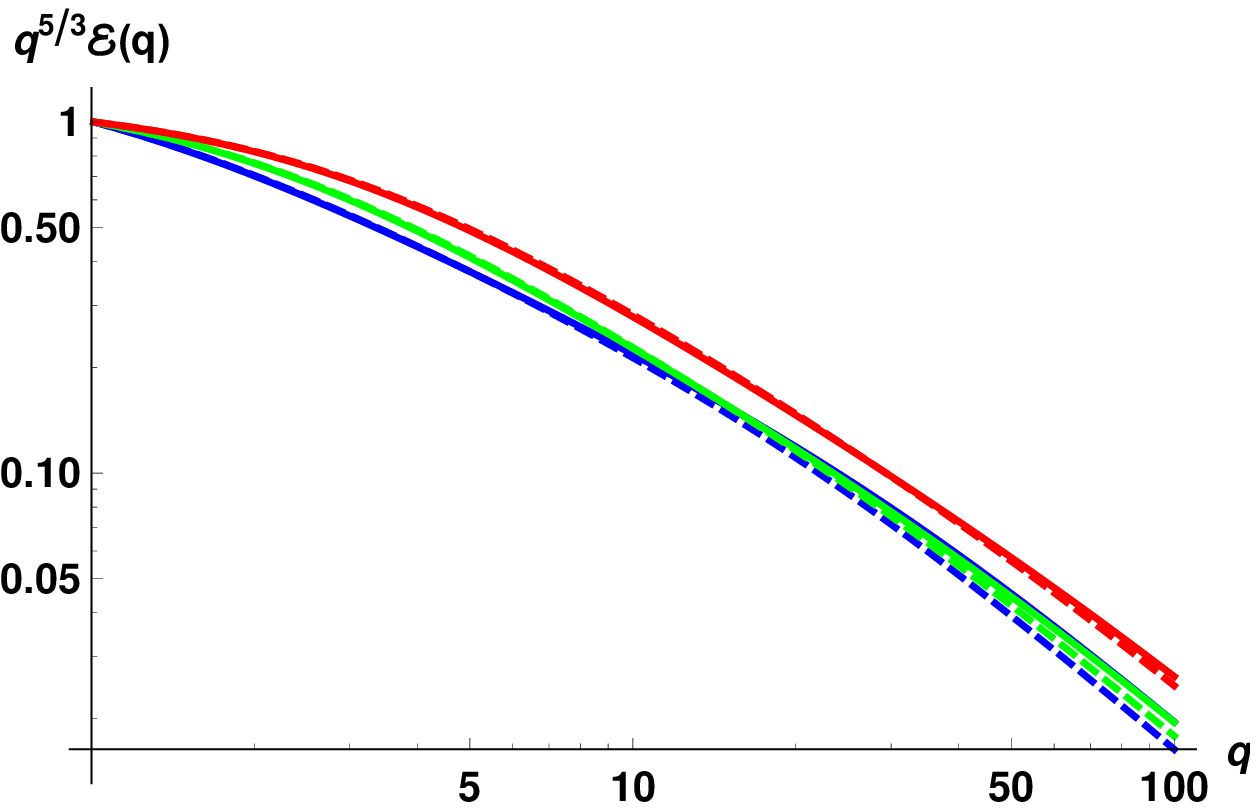}&
 		\includegraphics[scale=0.6]{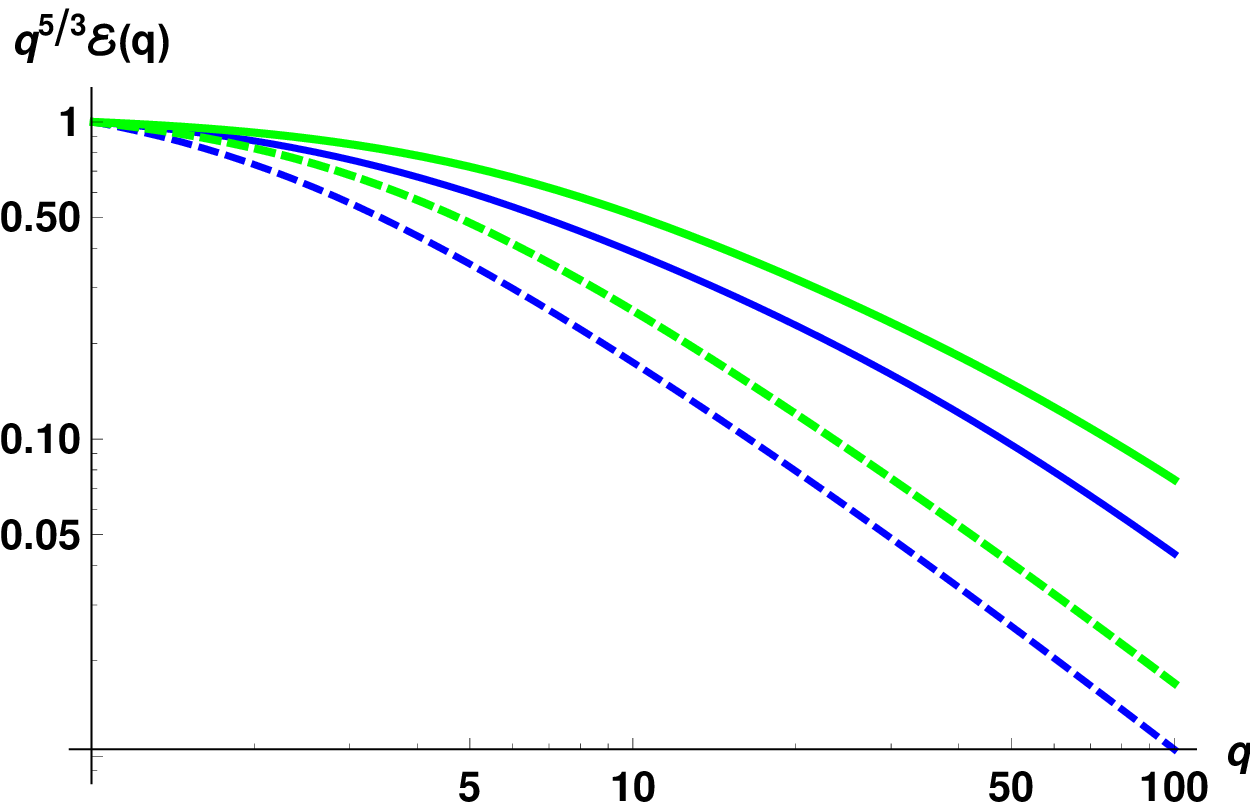}
 		\\
 	\end{tabular}
 	\caption{\label{ff:4} Color online. The compensated  energy spectra for experimental conditions at $T=1.65$K [Panel(a)], $T=1.85$K [Panel(b)], $T=2.0$K [Panel(c)] and $T=2.1\,$K Panel (d)]. The spectra for large heat fluxes $Q$ are shown by red lines [upper lines in panels (a)-(c)], for moderate heat fluxes-- by green lines and for small heat fluxes-- by lower blue lines. The  explicit values of $Q$ are given in Tab.\,\ref{t:2}. The normal fluid  spectra are shown by solid lines, the superfluid spectra -- by dashed lines. Note, that all spectra here are shown for  $q<100$, within  the applicability range of \Eqs{NSE} for the relevant flow parameters.  
 	}
\end{figure*}

The energy spectra for each of11 sets of measurements,  computed using \Eqs{20} with the corresponding parameters,  are collected  in \Fig{ff:4}.    At each temperature, the red lines
 correspond to the spectrum with the largest value of the heat flux $Q$, the green lines -- for the intermediate value of $Q$ and  the lowest  blue lines
 -- to the smallest  $Q$.  For these flow conditions, all spectra are strongly suppressed and are not scale-invariant, although the degree of the deviation from scale-invariance varies. Interestingly, at these conditions the normal fluid spectra for all temperatures, except $T=2.1\,$K, appear very similar to each other. To characterize the scaling behavior of these spectra, we use again \Eq{meanM} and calculate the mean exponents over  first decade $\<m \sb n\>\sb{10}$. 
 
\begin{figure}[t]
 	\includegraphics[width=0.45\textwidth]{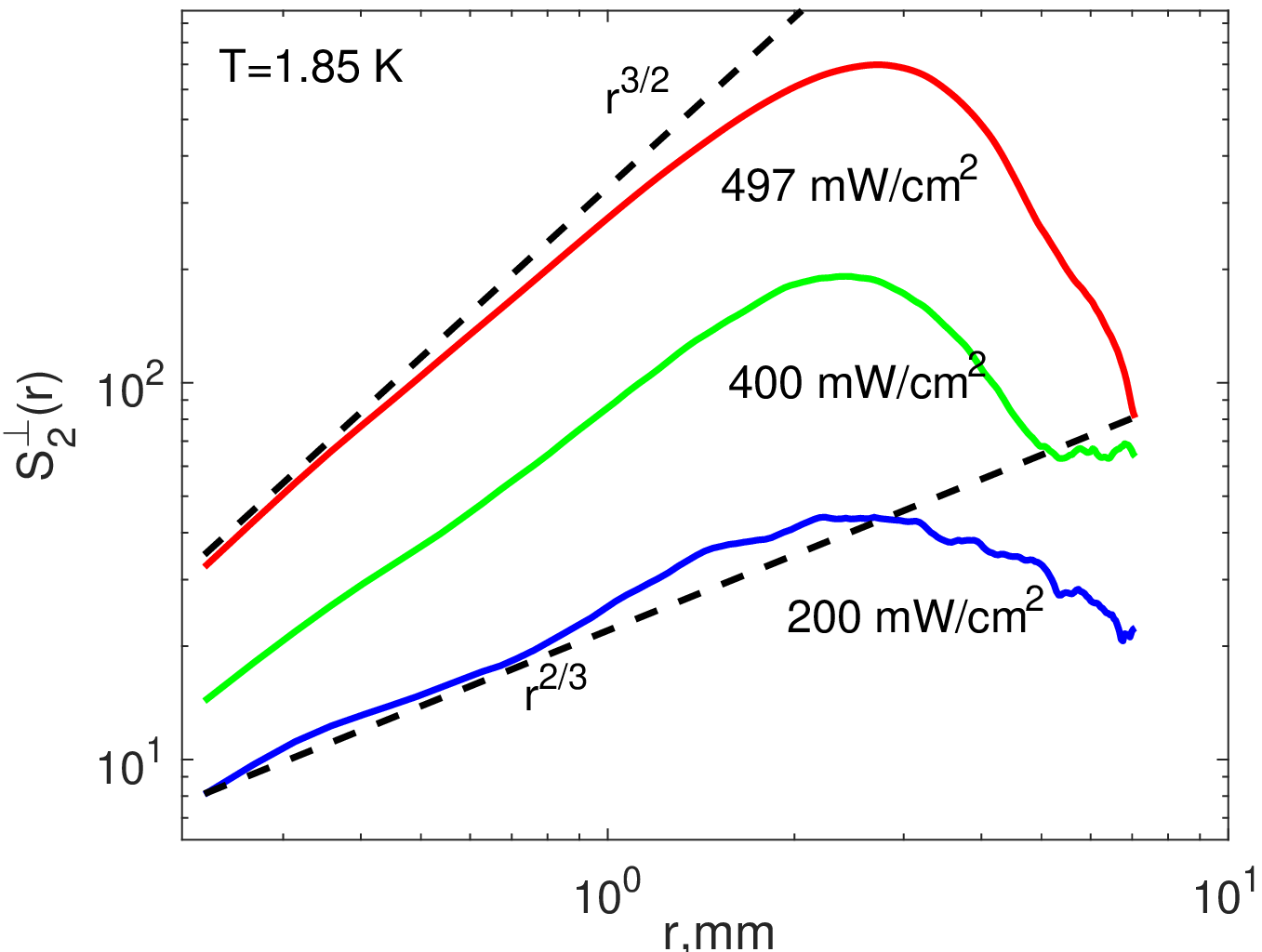}
 	\caption{\label{ff:7} Color online. The   2-nd order structure
 		functions\cite{WG-17b}  $S_2(r)$ as a function of separation  $r$  for
 		$T=1.85\,$K and different heat fluxes. The black dashed lines indicate
 		the scaling behavior $r^{2/3}$ and $r^{3/2}$ and serve to guide the eye only.}
 \end{figure}

The scaling behavior of such non-scale-invariant spectra do not have simple relation to the scaling exponents of the second order structure function.
 The experimentally measured 2nd order transversal structure functions $S^{\bot}_2(r)$  were found to exhibit a power-law behavior over an interval of scales of about one decade. The examples\cite{WG-17b} of the experimental $S^{\bot}_2(r)$  for $T=1.85\,$K and different heat fluxes are shown in \Fig{ff:7}. The scaling exponents $n\sb{exp}$ were measured by a linear fit over the corresponding  $r$-range  and it was suggested\cite{WG-17a} that the scaling exponents of the underlying energy spectra should scale as $n\sb{exp}+1$. We therefore compare the theoretical predictions $\<m \sb n\>\sb{10}$ with the  proposed experimental exponents $n\sb{exp}+1$, listed  in the columns \#12 and \#13,  Tab.\ref{t:2}. The error-bars for $n\sb{exp}+1$ correspond to the fit quality of $S^{\bot}_2(r)$. It was assumed\cite{WG-17a} that  additional   experimental inaccuracies (supposedly present in previous experiments\,\cite{WG-2015}) are absent.

 To rationalize the results, we first analyze the dimensionless model parameters, corresponding to  the experimental conditions. First of all, the dimensionless counterflow velocity $\~U\sb{ns}\approx 4$ for all conditions. Therefore, the differences in flow conditions for a given temperature are mostly translated to the differences in  efficiency of the dissipation by mutual friction $\~\Omega\sb{ns}$. As we saw in \Figs{ff:2} and \ref{ff:2b}, in the relevant range $\~\Omega\sb{ns}\simeq 3.5-26$,  the mean scaling exponents  $\<m\sb n\>\sb{10}$ are expected to be strongly affected by the dissipation due to mutual friction and to  weakly depend on $\~\Omega\sb{ns}$. Indeed, for relatively large $\~\Omega\sb{ns}$, the exponents $n\sb{exp}+1$ are clustered by temperatures and do not vary much. However, their values are smaller for $T=2.1\,$K than for $T=2.0\,$K, while exponents for $T=2\,$K and $T=1.85\,$K are compatible (except for the smallest heat flux, for which both the Re$\sb n$ and  $\~\Omega\sb{ns}$ are small). This trend agrees with the expected $T$ dependence, shown in  \Fig{ff:2}, if we also account for the difference in the Reynolds numbers (cf. \Fig{ff:3}). Remarkably, for these conditions the theoretical values of $\<m \sb n\>\sb{10}$  are in a good qualitative agreement with $n\sb{exp}+1$. Note that these values were obtained without any fitting parameters.
The discrepancy between experimental estimates and  theoretical predictions is limited to the flows with low Reynolds numbers ($T=1.65$K and $T=1.85$K, $Q=200$mW/cm$^2$). In these conditions, the flow inhomogeneity, not accounted for by our theory, may play important role. In particular, in the low Re channel flow, the width of the near-wall buffer layer is compatible\cite{Pope} with the width of the turbulent core (the region of well-developed turbulence around the channel centerline). Its  contribution to the velocity structure functions become significant. The typical size of the largest eddies in the buffer layer is not constant: it is of order of the local distance to the wall and smaller than the outer scale  in the turbulent core. A more accurate estimate of $\Delta$ (or $k_0$, which defines the dimensionless model parameters) for these conditions, may improve the agreement between the theory and experiment. For instance,  with all other parameters unchanged,  larger $k_0$ leads to smaller $\~\Omega\sb{ns}$. These smaller  $\~\Omega\sb{ns}$ correspond to the lower, fast-changing part of the $m$ vs $\~\Omega\sb{ns}$ curves in \Fig{ff:2b}. This kind of behavior is indeed observed at $T=1.65$K.  The simplifications of the theory in the description of the energy exchange between components are another possible reason for the discrepancy at low $T$. 
 
 \subsection*{\label{s:conc}Conclusions}
We developed a semi-quantitative theory of stationary, space-homogeneous isotropic developed counterflow turbulence in superfluid $^4$He.
 The theory captures basic physics of the energy spectra dependence on the main flow parameters and  accounts for the interplay between:\\

   (i) the turbulent velocity coupling by mutual friction,  dominant  at large scales $r>r_\times\simeq \pi/k_\times$;
   
   (ii) its decoupling, caused by the sweeping of the normal and superfluid eddies in the opposite directions, which becomes important at scales  $r<r_\times$; 
   
   (iii) the  turbulent energy dissipation due to mutual friction at scales  $r<r_\times$, that gradually decreases the energy flux over scales and suppresses the energy spectrum, similar to the turbulence in $^3$He.

   The  ultra-quantum peak, well separated from the   quasi-classical interval of scales $r>\ell$,  serves in our theory as a space- and time-independent source of the vortex-line density $\C L$ involved in the mutual frequency force,   $\propto \C L(\B u\sb n - \B u\sb s)$.

The resulting energy spectra of the normal and superfluid components are greatly suppressed with respect to their classical fluid counterpart. Moreover, the spectra are non-scale invariant,  strongly depend on the temperature and the counterflow velocity.  Their scaling behavior may be characterized by local
slopes. These slopes, calculated at the largest scales (smallest wavenumbers) depend not-monotonically on the mutual-friction frequency. The deviation from  scale-invariance is evident by comparison of the outer-scale slope $m_j(1)$ with the mean over an interval slope $\<m_j\>_q$. The small scale behavior is further affected by the viscous dissipation. This effect is most prominent for the normal spectra at high $T$ and for the superfluid spectra at low $T$. 

By comparing the mean scaling exponents, calculated over the interval $k\in [k_0-10k_0]$ without any fitting parameters, with the experimental estimates $n\sb{exp}+1$, we find a  good qualitative agreement between our theory and observations for $T\gtrsim 1.85$K. This allows us to believe that most important  simplifications used  in developing  the theory:
 
 i) the space homogeneity and isotropy of the flow;
 
 ii) the uncontrolled approximations in the derivations of the differential closure and the decorrelation function;
 
 iii)  further simplification of the cross-correlation function that ignores the energy flux between the normal and superfluid subsystems,

play just a secondary role and may be relaxed in later development.  In particular,  the energy transfer between the fluid components by mutual friction force is expected to affect the scaling behavior of both spectra, especially at low $T$. Therefore a better approximation for the  cross-correlation function   may account for this effect. The  possible influence of the flow space-inhomogeneity and anisotropy  may be responsible for the differences between the apparent scaling behavior of the transverse structure functions and of the isotropic 3D energy spectra. An account for these factors is beyond the scope of this paper.

 \acknowledgements We are grateful to W. Guo (Florida State University, FL,
 USA) for productive discussions.


\begin{thebibliography}{99}
\bibitem{CK}Note that the original K41 constant enters a relation $E(k)=C\sb K \langle \epsilon \rangle^
 	{2/3}k^{5/3}$.  In \Eq{K41} we define the energy flux via the energy spectum and therefore the constant $C=C^{-3/2}\sb K$.
 	In this work we adopt the value of $C\sb K=1.62$ obtained by Yeung and Zhou\cite{YeungZhou}.
\end{thebibliography}
\end{document}